\documentclass[11pt]{article}

\usepackage[round]{natbib}
\usepackage{fullpage}
\usepackage[final]{hyperref}
\usepackage{url}
\usepackage{algorithm}
\usepackage{algorithmic}
\usepackage{boxedminipage}
\usepackage{comment}
\usepackage{eqparbox}
\usepackage{multirow}
\usepackage{caption}

\usepackage{amsmath}
\usepackage{amsfonts}
\usepackage{graphicx}

\newlength{\commentindent}
\newcommand\matvec{{\it matvec}}

\usepackage{array}
\newcolumntype{R}[1]{>{\raggedleft\let\newline\\\arraybackslash\hspace{0pt}}m{#1}}

\newcommand\BibTeX{{\rmfamily B\kern-.05em \textsc{i\kern-.025em b}\kern-.08em
T\kern-.1667em\lower.7ex\hbox{E}\kern-.125emX}}

\newcommand{\kernel}[1]{\vspace{4pt}\newline\noindent{\bf{\emph{Kernel #1}}}:}
\newcommand{\kernelfl}[1]{\vspace{4pt}\noindent{\bf{\emph{Kernel #1}}}:}
\begin{document}


\title{Acceleration of tensor-product operations for high-order finite element methods.}

\author{Kasia \'{S}wirydowicz\thanks{Department of Mathematics, Virginia Tech, McBryde Hall,
24061 Blacksburg, VA,  USA, kswirydo@vt.edu}, Noel Chalmers, Ali Karakus and T. Warburton}


\maketitle

\begin{abstract}
This paper is devoted to GPU kernel optimization and performance analysis of three tensor-product operators arising in finite element methods. We provide a mathematical background to these operations and implementation details. Achieving close-to-the-peak performance for these operators requires extensive optimization because of the operators' properties: low arithmetic intensity, tiered structure, and the need to store intermediate results inside the kernel. We give a guided overview of optimization strategies and we present a performance model that allows us to compare the efficacy of these optimizations against an empirically calibrated roofline. 
\end{abstract}

\tableofcontents

\section{Introduction}

State-of-the-art high-order finite element method (FEM) based simulation codes now execute calculations on parallel machines with up to a million CPU cores. For example, the Nek5000 incompressible fluid flow simulator has successfully  scaled up to a million CPU cores using an MPI distributed computing model (\citep{Fischer2015b}). The current shift towards greater on-chip parallelism makes it imperative to focus on finer grain  parallel implementation of high-order operators on next generation accelerators, such as Graphics Processing Units (GPUs). Efficient high-order finite element  implementations rely heavily on tensor-product contractions (\citep{deville2002high}). These elemental tensor-product operations are memory bound and require repeated sweeps through the data for each finite element operation. In this work we demonstrate that despite their complexity, it is possible to achieve nearly maximum throughput on a current generation GPU.

In this paper we focus in particular on GPU implementations of three common place matrix-vector multiplications (further abbreviated as \textit{matvecs}) arising in finite element codes: mass \matvec{} (BP1.0), stiffness \matvec{}  (BP3.5), and stiffness \matvec{} that involves de-aliasing integration (BP3.0). These problems have been formulated as a part of CEED co-design effort (\citep{ceed2017}). BP1.0, BP3.5, and BP3.0  appear to be potentially well-suited for fine-grain parallelism but they are challenging to fully optimize. These benchmarks are characterized by low arithmetic intensity, namely that the ratio of floating point operations (FLOPS) to data transfer is low. The benchmarks also require non-unitarily strided memory access patterns since all threads that are processing a single element need the data computed by other threads that process this element. Finally, the operations consist of several concatenated tensor-product contractions and all threads simultaneously process the contractions in a  prescribed order.

We detail several strategies that are collectively applied to maximize the computational throughput of these operations. In a sequence of computational experiments we progressively optimize the operations by varying the amount of global memory accesses, shared memory usage, register variable usage, data padding, and loop unrolling. 

To guide the optimization process we construct a performance model that serves as a tool to evaluate the efficiency of the kernels. Because of the nature of the finite element operations, the performance of our kernels is limited by the bandwidth of the data streaming from both the global device memory and shared memory as well as finite register file and shared memory capacity. While a theoretical performance roofline, as presented by \citep{Lo2015}, gives an upper bound on the capability of the hardware under ideal circumstances, we instead rely on an empirically calibrated roofline model that provides a realistic estimate of the best achievable performance.

\subsection{Overview of published literature}

Early optimization studies of FEM operations date back to 2005 when \citep{Goddeke2005} and  \citep{Goddeke2007}  solved elliptic PDEs on two-dimensional domains using a cluster equipped with GPU accelerators. In later work, \citep{Goddeke2009} focused mostly on accelerating the Navier-Stokes solver for lid driven cavity and laminar flow over a cylinder. The study presented by \citep{fu2014} targeted the entire FEM pipeline using the elliptic Helmholtz problem to show how the FEM codes are ported to the GPU. In addition, \citep{fu2014} discussed strategies for accelerating conjugate gradient and algebraic multigrid solvers.

Other authors usually choose to work on improving the performance of one part of the FEM pipeline. For example, \citep{cecka2011} and \citep{Markall2013} focused on the global assembly phase of FEM and showed how to optimize this phase for GPU execution. Markall and co-authors emphasize that the choice of the most efficient algorithm strongly depends on the available hardware and selected programming model. In their work, Markall and co-authors considered the CPU and the GPU with OpenCL and CUDA parallel implementations. Moreover, \citep{markall2010, Markall2013} argue that making the code portable between different threading systems (such as many-core and multi-core architectures) requires a high-level programming language. 

In all of the above work the greatest concern was efficiently pipelining GPU execution of FEM with low-order discretizations. In contrast, \citep{Remacle2016} present algorithms for solving elliptic equations using high-order continuous hexahedral finite elements on the GPUs. The issues associated with efficiently handling the greater complexity of high-order FEM operations highlighted in that work are amplified in the current paper and refined with the use of a new roofline model.


The evaluation of the \matvec{} typically accounts for the highest cost of the elliptic FEM solver \citep[Table 2, Table 3]{Remacle2016}. Optimization of \matvec{} performance on the GPUs appears in the literature mostly in the context of general applications. Yet a few papers focused directly on the low-order FEM \matvec{}. For example,~\citep{Dziekonski2017} used a conjugate gradient solver and optimized \matvec{} product as its important part. \citep{Dehnavi2010} and \citep{Grigoras2016} present similar findings and provided optimization strategies. However, in our case the action of the operator is local to the element. We never assemble the global matrix and we perform the element-wise \matvec{}. In the literature this approach is known as \textit{matrix-free} approach.

The \textit{matvecs} used in this paper can be expressed as a concatenation of tensor contractions. Related work on efficient implementation of tensor contractions for the GPUs consists of several papers. \citep{Nelson2015} applied high level language to formulate tensor contractions and used GPU code auto-tuning to enhance the performance of the code. The findings were tested using several benchmark problems, including a problem derived from Nek5000 (see \citep{nek5000}). Optimized code reached 120 GFLOPS/s on a Maxwell class GPU. \citep{Abdelfattah2016} published similar work where, in addition to specifying high-level language to formulate tensor contractions, the authors introduced a performance model based on both the number of floating point operations and the number of bytes of global memory that need to be read and written. The performance results were compared to CUBLAS and to CPU code executed using 16 CPU cores. The most efficient version of the code reaches 180 GFLOPS/s on NVIDIA Tesla K40 GPU. While two previously mentioned papers focused on small tensor contractions, \citep{Liu2017} optimized large tensor contractions, also using auto-tuning for performance optimization. The paper explained the details of the implementation and reported the global memory usage.

The performance model in this paper is rooted in earlier work by several authors, originating from the {\it roofline model} in~\citep{Lo2015}. \citep{Stratton2012} and \citep{Zhang2011} who modeled the GPU performance using benchmarks. These efforts focused on a semi-automatic way of identifying performance bottlenecks. Acquiring a more complete understanding of the performance of the code on the GPU and designing benchmark specifically to reveal the underlying hardware properties frequently appears in the literature due to lack of available complete GPU hardware documentation. At the same time, hardware organization strongly impacts the performance (~\citep{Wong2010}, \citep{Hong2009}, and \citep{Lee2010}). \citep{Volkov2008} and~\citep{volkov-2010} employed a different approach; they optimized matrix-matrix multiplication on the GPU and aimed to explain the code performance while improving the implementation. In his 2010 paper, Volkov pointed out that, contrary to a common belief, high GPU occupancy is not necessary to obtain close the peak performance declared by the GPU manufacturer. This observation is important for the current paper. Our most efficient kernels are characterized by high usage of registers and shared memory, thus the achieved occupancy can often be as low as $10\%$ even though the performance of these kernels is close to the empirical roofline. Volkov observation that the aggregate number of shared memory accesses per thread should be reduced to avoid bottlenecks motivates a final optimization for the high-order FEM operations considered in this work. We exploit the intrinsic structure of the interpolation and differentiation tensors used in the construction of the action of the mass and stiffness matrices to improve throughput by reducing the aggregate number of shared memory accesses.

The remainder of this paper is organized as follows. We first present the mathematical description of the three benchmark problems mentioned earlier. We then introduce the performance model we use to evaluate our kernels. We continue by explaining the implementation and optimization choices for each of the benchmark problems. Finally, we gives some concluding remarks and future goals.

\section{Notation}

\begin{table*}[t]
 \begin{tabular}{>{\centering}m{0.12\textwidth}|>{\centering}m{0.12\textwidth}|p{0.65\textwidth}} \hline
 \multicolumn{2}{c|}{Symbol} & \\
  \multicolumn{1}{c}{\bf Code} &  \multicolumn{1}{c|}{\bf Math} & {\bf Meaning }\\
  \hline
 \texttt{N} & $N$ & Degree of the polynomial used in the interpolation\\
 \texttt{Nq} & $N_q$ & Number of Gauss-Lobatto-Legendre (GLL) quadrature nodes: $N_q = N+1$ \\
 \texttt{Np} & $N_p$ &  Number of GLL nodes in a hexahedral element: $N_p = N_q^3$ \\
\texttt{glNq} & $N_q^{GL}$ &  Number of Gauss-Legendre (GL) quadrature nodes:  $N_q^{GL} = N+2$ \\
\texttt{glNp} & $N_p^{GL}$ & Number of GL nodes in a hexahedral element. $N_p^{GL} =\left(N_q^{GL}\right)^3$\\
\texttt{I} & $\mathbf{I}^{\mathrm{1D}}$ & $(N+2) \times (N+1)$ interpolation matrix from GLL nodes to GL nodes \\
\texttt{D} & $\mathbf{D}^{\mathrm{1D}}$ & $(N+1) \times (N+1)$ differentiation matrix used in BP3.5, see (\ref{eq:derivativeOps})\\
\texttt{glD} & $\tilde{\mathbf{D}}^{\mathrm{1D}}$ & $(N+2) \times (N+2)$ differentiation matrix, used in BP3.0\\
\texttt{NElements} & $N_{el}$ & Number of elements in the mesh \\ \hline
 \end{tabular}
 \caption{The notation used in in the paper. The first column shows the symbols used in pseudocode listings and the second column shows the symbols used in the derivations. \label{tab:notation}}
 \end{table*}

The notation used throughout this paper is shown in Table~\ref{tab:notation}. In addition, we define $$
1\, \mbox{TFLOPS} = 10^{12}\, \mbox{FLOPS}
$$
The floating point throughput rates for compute kernels in this paper are reported using TFLOPS/s. We use double precision arithmetic in all the tests.

\section{Hardware and software} \label{hardware.sec}

All computational studies in this paper were performed on a single Tesla P100 (Pascal class) PCI-E GPU with 12 GB RAM and maximum theoretical bandwidth of 549 GB/s. The GPU is hosted by a server node equipped with Intel Xeon E5-2680 v4 processor, 2.40 Ghz with 14 cores. The code was compiled using the gcc 5.2.0 compiler and the nvcc CUDA compiler release 8.0, V8.0.61 managed by the OCCA library (see \citep{Medina2014}).

We compute the reference times (needed for copy bandwidth in roofline plots) using CUDA events. The kernels are timed using \verb|MPI_Wtime|.

In our experiments, we test the code on two hexahedral cube-shaped meshes. The small mesh consists of 512 elements and the large mesh consists of 4096 elements.

\section{Problem description}

The three benchmark problems we consider below were formulated by the Center for Efficient Exascale Discretizations (see \citep{ceed2017}). These problems are motivated by considering the numerical approximation of the screened Poisson equation
\begin{equation}\label{eq:screenedPoisson}
-\nabla^2 u + \lambda u = f,    
\end{equation}
 by a high-order finite element method on hexahedral elements. In the equation, $f$ is a given forcing function and $\lambda$ is a constant. To begin we consider an unstructured mesh of a domain $\mathcal{D} \subset \mathbb{R}^3$ into $K$ hexahedral elements $D^e$, where $e=1,\ldots,K$, such that
\begin{equation}
\label{eq:elSum}
\mathcal{D} = \bigcup_{e=1}^{K} D^e.
\end{equation}
We assume that each hexahedral element $D^e$ with vertices $\{x_n,y_n,z_n\}_{n=1}^{n=8}$ is the image of the reference bi-unit cube $\hat{D}$ under a tri-linear map. We take the reference cube to be the bi-unit cube, i.e.
\[
\hat{D} = \{(r,s,t): -1 \leq r,s,t \leq1\}.
\]
On each element we consider the variational form of the screened Poisson problem \eqref{eq:screenedPoisson} on element $D^e$ by requiring that $u$ satisfies 
\[
\int_{D^e} \nabla v \cdot \nabla u \; dV  +  \lambda \int_{D^e} v u \; dV = \int_{D^e} v f \; dV.
\]
for all test functions $v \in H^1(D^e)$. We map the integrals to the reference cube $\hat{D}$ in order to write the variational form as
\[
\int_{\hat{D}} \nabla v^T G^e \nabla u^e \; dV  +  \lambda \int_{\hat{D}} v u^e |J^e|\; dV = \int_{\hat{D}} v f |J^e|\; dV.
\]
where $|J^e|$ is the determinant of the Jacobian $J^e$ of the mapping from element $D^e$ to the reference cube $\hat{D}$ and $G^e = |J^e| (J^e)^T J^e$ is the scaled elemental metric tensor. The $\nabla$ operators in these integrals are now understood to be in $(r,s,t)$-space.

On the reference cube $\hat{D}$ we construct a high-order finite element approximation of the function $u$, denoted $u^e$, which is a degree $N$ polynomial in each dimension. We denote the space of all such polynomials as $\mathbb{Q}^N(\hat{D})$. As a polynomial basis of $\mathbb{Q}^N(\hat{D})$ we choose the tensor product of one-dimensional Lagrange interpolation polynomials $\{l_i\}_{i=0}^{i=N}$ based on $N+1$ Gauss-Lobatto-Legendre (GLL) nodes on the interval $[-1,1]$. We denote the GLL nodes in $[-1,1]$ by $\{r_i\}_{i=0}^{i=N}$, and use an analogous notation for the GLL nodes in the $s$ and $t$ dimensions. Using a multi-index, we define the multi-dimensional basis polynomials $l_{ijk}(r,s,t)$ to be the tensor product of the one-dimensional Lagrange interpolating polynomials, i.e.
\[
l_{ijk}(r,s,t) = l_i(r)l_j(s)l_k(t),
\]
for all $0\leq i,j,k \leq N$. Hence, on $\hat{D}$ we can express the polynomial $u^e$ as 
\[
u^{e}\left(r,s,t\right) = \sum_{i=0}^{N} \sum_{j=0}^N \sum_{k=0}^N u^e_{ijk} l_{ijk}\left(r,s,t\right).
\]
Consequently, the coefficients $u_{ijk}^e$ are the nodal values of the polynomial $u^e$ at the GLL interpolation points $(r_i,s_j,t_k)$, i.e., $u_{ijk}^e = u^e(r_i,s_j,t_k)$. 

Taking the test functions $v$ to be each of the basis Lagrange interpolating polynomials, $v=l_{i'j'k'}(r,s,t)$, and taking the vector $\mathbf{u}^e$ to be the vector of the polynomial coefficients of $u^e$, i.e. $\mathbf{u}^e = [u^e_{000}, u^e_{001},\ldots, u^e_{NNN}]^T$, we can write the variational formulation as the following linear system
\[
\mathbf{S}^e \mathbf{u}^e + \lambda \mathbf{M}^e \mathbf{u}^e = \mathbf{f}^e
\]
where the local stiffness matrix $\mathbf{S}^e$, local mass matrix $\mathbf{M}^e$, and local load vector $\mathbf{f}^e$ are defined as
\begin{align}
    \mathbf{S}^e_{ijk,i'j'k'} &= \int_{\hat{D}} (\nabla l_{i'j'k'})^T G^e \nabla l_{ijk} \; dV, \label{eq:stiffnessDef}\\
    \mathbf{M}^e_{ijk,i'j'k'} &= \int_{\hat{D}} l_{i'j'k'} l_{ijk} |J^e| \; dV, \label{eq:massDef}\\
    \mathbf{f}^e_{i'j'k'} &= \int_{\hat{D}}  l_{i'j'k'} f |J^e| \; dV.
\end{align}

We concatenate the local stiffness and mass matrix operators as well as local load vectors to form global unassembled versions which we denote, $\mathbf{S}$, $\mathbf{M}$ and $\mathbf{f}$, respectively. Doing so, we form the following a block diagonal system which operators on the global vector of solution coefficients, which we denote $\mathbf{u}$, 
\begin{equation*}
    \mathbf{S} \mathbf{u} + \lambda \mathbf{M} \mathbf{u} = \mathbf{f}.
\end{equation*}
Due to its block diagonal structure, these global stiffness and mass matrix operators can be applied in a matrix-free (element-wise) way and no communication is required between elements.

The benchmarks presented below consider the efficient action of either just the mass matrix $\mathbf{M}$ or the complete screened Poisson operator $\mathbf{S} + \lambda\mathbf{M}$ on each element. Benchmark problem 1.0 and 3.0 consider the case where the integrals \eqref{eq:stiffnessDef}-\eqref{eq:massDef} are evaluated using a full Gauss-Legendre (GL) quadrature and benchmark problem 3.5 considers the case where the integrals are approximated using simply a quadrature at the GLL interpolation points.

\label{theory.sec}
\subsection{Benchmark Problem 1: Mass Matrix Multiplication}
\label{theory.BP1.subsec}
The first benchmark we present involves matrix-vector product of the high-order finite element mass matrix $\mathbf{M}$ and a corresponding vector. This operation is a component of finite element elliptic operators and some preconditioning strategies. Thus, it serves as a useful initial performance benchmark. The operation requires relatively little data transfers compared with more demanding differential operators which necessitate loading the elemental metric tensor for each node of the element, as discussed in later benchmarks.

Beginning from the integral definition of the entries of the local mass matrix $\mathbf{M}^e$ in \eqref{eq:massDef} we use the fact that $\hat{D}$ is the reference cube to evaluate the integral in each dimension separately via an $N_q^{GL}$-node Gauss-Legendre (GL) quadrature. We denote the quadrature weights and nodes in the $r$ dimension as $\{\tilde{w}_a\}_{a=1}^{a=N_q^{GL}}$ and $\{\tilde{r}_a\}_{a=1}^{a=N_q^{GL}}$, respectively, and use an analogous notation for quadrature nodes in the $s$ and $t$ dimensions. Thus we can write
\begin{equation}
\mathbf{M}^e_{ijk,i'j'k'} = \sum_{a=1}^{N_q^{GL}} \sum_{b=1}^{N_q^{GL}} \sum_{c=1}^{N_q^{GL}} \tilde{w}_a \tilde{w}_b \tilde{w}_c |J^{e}(\tilde{r}_a,\tilde{s}_b,\tilde{t}_c)|  l_{i'j'k'}(\tilde{r}_a,\tilde{s}_b,\tilde{t}_c)l_{ijk}(\tilde{r}_a,\tilde{s}_b,\tilde{t}_c). \label{eq:massQuad}
\end{equation}
We can write this expression in matrix form by defining the interpolation operator,
\[
\mathbf{I}_{abc,ijk} = l_{ijk}(\tilde{r}_a,\tilde{s}_b,\tilde{t}_c),
\]
for all $i,j,k=0,\ldots,N$ and $a,b,c=1,\ldots,N_q^{GL}$, as well as the diagonal matrix $\mathbf{J}^e$ of weights and geometric data which gas entries
\[
\mathbf{J}^e_{abc,abc} = \tilde{w}_a \tilde{w}_b \tilde{w}_c |J^{e}(\tilde{r}_a,\tilde{s}_b,\tilde{t}_c)|,
\]
for all $a,b,c=1,\ldots,N_q^{GL}$ in order to write the local mass matrix \eqref{eq:massQuad} compactly as
\[
\mathbf{M}^e = \mathbf{I}^T \mathbf{J}^e \mathbf{I}.
\]
Note that since the basis interpolation polynomials are a tensor products of one-dimensional polynomials, we can define the one-dimensional interpolation matrix $\mathbf{I}^{1D}$ as
\begin{equation}
\mathbf{I}^{\mathrm{1D}}_{ai} = l_i(\tilde{r}_a), \label{eq:interpolationOps} \\
\end{equation}
for all $i=0,\ldots,N$ and $a=1,\ldots,N_q^{GL}$, in order to express the interpolation operator $\mathbf{I}$ as a tensor product of the one-dimensional operators, i.e. 
\[
\mathbf{I} = \mathbf{I}^{\mathrm{1D}}\otimes \mathbf{I}^{\mathrm{1D}}\otimes\mathbf{I}^{\mathrm{1D}}.
\]
Thus, the interpolation operation from the GLL interpolation nodes to the GL quadrature nodes can be applied using three tensor contractions, while the projection back to the GLL nodes via the transpose interpolation operator comprises three additional tensor contractions. Since the remaining operation is simply the multiplication by the diagonal matrix $\mathbf{J}^e$ no further tensor contractions are required. Since we will make use of the interpolation and projection operations again in BP3.0 below, we detail their pseudo code in Algorithms \ref{alg:interp} and \ref{alg:proj}. We then detail the full matrix free action of the mass matrix in the pseudo code in Algorithm \ref{alg:bp1alg}.

\begin{algorithm*}[ht]
  \caption{Interpolation from GLL to GL nodes}
  \label{alg:interp}
\begin{boxedminipage}{\textwidth}
    \begin{algorithmic}[1]
    \STATE {\bf Data:} (1) $\mathbf{q}^{e}$, size  $N_p$; (2) Interpolation matrix $\mathbf{I}^{\mathrm{1D}}$, size $N^{GL}_q\times N_q$\; 
    \STATE {{\bf Output: } $\tilde{\mathbf{q}}^e$, size  $N_p^{GL}$}\;
    \FOR {$c,a\in \left\{1, \ldots N_q\right\}, \;j \in \left\{1, \ldots, N_q^{GL}\right\}$ }
        \STATE $\hat{\mathbf{q}}_{cja}^{e}=\sum_{b=1}^{N_q}\mathbf{I}_{jb}^{\mathrm{1D}} \mathbf{q}_{cba}^{e}$\COMMENT{Interpolate in $b$ direction}\;
    \ENDFOR
     \FOR {$c \in \{1,2, \ldots N_q\}, \; i,j \in\{1,2 \ldots, N_q^{GL}\}$ }
        \STATE $\hat{\hat{\mathbf{q}}}_{cja}^{e}=\sum_{a=1}^{N_q}\mathbf{I}_{ia}^{\mathrm{1D}} \tilde{\mathbf{q}}_{cja}^{e}$\COMMENT{Interpolate in $a$ direction}
    \ENDFOR
     \FOR {$k, i,j \in\{1,2 \ldots, N_q^{GL}\}$ }
        \STATE $\tilde{\mathbf{q}}_{kji}^{e}=\mathbf{J}_{kji}^{e} \sum_{c=1}^{N_q}\mathbf{I}_{kc}^{\mathrm{1D}} \hat{\hat{\mathbf{q}}}_{cji}^{e}$\COMMENT{Interpolate in $c$ direction, save}
    \ENDFOR
  \end{algorithmic}
\end{boxedminipage}
\end{algorithm*}

\begin{algorithm*}[ht]
  \caption{Projection from GL to GLL nodes}
  \label{alg:proj}
\begin{boxedminipage}{\textwidth}
    \begin{algorithmic}[1]
    \STATE {\bf Data:} (1) $\tilde{\mathbf{q}}^e$, size  $N_p^{GL}$; (2) Interpolation matrix $\mathbf{I}^{\mathrm{1D}}$, size $N^{GL}_q\times N_q$\; 
    \STATE {{\bf Output: } $\mathbf{q}^e$, size  $N_p$}\;
    \FOR {$k, i \in\{1,2 \ldots, N_q^{GL}\},\; b \in\{1,2 \ldots, N_q\}$ }
        \STATE $\hat{\hat{\mathbf{q}}}_{kbi}^{e} = \sum_{j=1}^{N_q^{GL}} \mathbf{I}_{jb}^{\mathrm{1D}}\tilde{\mathbf{q}}_{kji}^{e}$\COMMENT{Project in $b$ direction}
    \ENDFOR
     \FOR {$k \in\{1,2 \ldots, N_q^{GL}\}\; b, a \in\{1,2 \ldots, N_q\}$ }
        \STATE $\hat{\mathbf{q}}_{kja}^{e}= \sum_{i=1}^{N_q^{GL}} \mathbf{I}_{ia}^{\mathrm{1D}}\hat{\hat{\mathbf{q}}}_{kbi}$\COMMENT{Project in $a$ direction}
    \ENDFOR
    \FOR {$ c,b,a  \in\{1,2 \ldots, N_q\}$ }
        \STATE $\mathbf{q}_{cba}^{e}= \sum_{k=1}^{N_q^{GL}} \mathbf{I}_{kc}^{\mathrm{1D}}\hat{\mathbf{q}}_{kba}$\COMMENT{Project in $c$ direction, save}
    \ENDFOR
  \end{algorithmic}
\end{boxedminipage}
\end{algorithm*}

\begin{algorithm*}[ht]
  \caption{BP1.0: mass matrix multiplication}
  \label{alg:bp1alg}
\begin{boxedminipage}{\textwidth}
    \begin{algorithmic}[1]
    \STATE {\bf Data:} (1) $\mathbf{q}$, size  $N_{el} \cdot N_p$; (2) Interpolation matrix $\mathbf{I}^{\mathrm{1D}}$, size $N^{GL}_q\times N_q$; (3) Scaled Jacobians, $\mathbf{J}$, size $N_{el}\times N_p^{GL}$\; 
    \STATE {{\bf Output: } $\mathbf{Mq}$, size  $N_{el}\cdot N_p$}\;
    \FOR {$e\in\left\{1,2, \ldots N_{el}\right\}$}
    \STATE $\tilde{\mathbf{q}}^{e}= \mathrm{Interpolate}(\mathbf{q}^e,\mathbf{I}^{\mathrm{1D}})$ \COMMENT{Interpolate to GL nodes (Algorithm \ref{alg:interp})}\;
     \FOR {$k, i,j \in\{1,2 \ldots, N_q^{GL}\}$ }
        \STATE $\tilde{\mathbf{q}}_{kji}^{e}=\mathbf{J}_{kji}^{e}\tilde{\mathbf{q}}_{kji}^{e}$  \COMMENT{Scale by Jacobian and integration weights}
    \ENDFOR
    \STATE $\mathbf{q}^{e}= \mathrm{Project}(\tilde{\mathbf{q}}^e,\mathbf{I}^{\mathrm{1D}})$ \COMMENT{Project to GLL nodes (Algorithm \ref{alg:proj})}\;
    \ENDFOR
  \end{algorithmic}
\end{boxedminipage}
\end{algorithm*}

\subsection{Benchmark Problem 3.5: Stiffness Matrix with Collocation Differentiation}
\label{theory.BP35.subsec}
For our second benchmark, we consider the matrix-vector product of the full high-order finite element screened Poisson operator $\mathbf{S} + \lambda\mathbf{M}$ and a corresponding vector. In this benchmark, the operators $\mathbf{S}$ and $\mathbf{M}$ are evaluated using a collocation GLL quadrature rather than the more accurate GL quadrature used in the other two benchmark problems. This operation is central to many elliptic finite element codes and is usually a part of a discrete operator we wish to invert. For example, incompressible flow solvers such as Nek5000 (see \citep{nek5000}) require solving a Poisson potential problem at each time step. Consequently, this \matvec{} is potentially evaluated many times in each time step of a flow simulation, making its optimization a significant factor for good performance. 

To describe the application of the full screened Poisson operator we begin by describing the local stiffness matrix $\mathbf{S}^e$ defined in \eqref{eq:stiffnessDef}. We evaluate the integral in \eqref{eq:stiffnessDef} in each dimension separately, this time using the $N+1$ GLL interpolation nodes as the quadrature. We denote the GLL quadrature weights and nodes in the $r$ dimension as $\{w_a\}_{a=0}^{a=N}$ and $\{r_a\}_{a=0}^{a=N}$, respectively, and use an analogous notation for the GLL quadrature nodes in the $s$ and $t$ dimensions. Thus we can write
\begin{equation}
\mathbf{S}^e_{ijk,i'j'k'} = \sum_{a=0}^{N} \sum_{b=0}^{N} \sum_{c=0}^{N} w_a w_b w_c (\nabla l_{i'j'k'}(r_a,s_b,t_c))^T G^{e}(r_a,s_b,t_c) \nabla l_{ijk}(r_a,s_b,t_c). \label{eq:stiffQuad}
\end{equation}
To write this expression in a more compact matrix form we begin by defining the differentiation operators,
\begin{align}
\mathbf{D}^r_{abc,ijk} &= \frac{\partial l_{ijk}}{\partial r} (r_a,s_b,t_c), \nonumber\\
\mathbf{D}^s_{abc,ijk} &= \frac{\partial l_{ijk}}{\partial s} (r_a,s_b,t_c), \label{eq:derivativeOps}\\
\mathbf{D}^t_{abc,ijk} &= \frac{\partial l_{ijk}}{\partial t} (r_a,s_b,t_c), \nonumber
\end{align}
for $i,j,k, a,b,c = 0,\ldots N$. We then define the gradient operator $\mathbf{D}$ to be the vector of these three derivative operators, i.e. $\mathbf{D} = [\mathbf{D}^r,\mathbf{D}^s,\mathbf{D}^t]^T$. Next, since $G^e$ is the scaled metric tensor on element $D^e$ defined by $G^e = |J^{e}|(J^e)^T J^e$, we denote the entries of $G^e$ as
\[
G^{e} =\left(\begin{matrix}
G^e_{rr} & G^e_{rs} & G^e_{rt} \\
G^e_{sr} & G^e_{ss} & G^e_{st} \\
G^e_{tr} & G^e_{ts} & G^e_{tt} 
\end{matrix}\right),
\]
We define a matrix $\mathbf{G}^e$ of operators where each entry of $\mathbf{G}^{e}$ is a diagonal matrix of weights and geometric data from $G^e$. That is, we define
\begin{equation}
\mathbf{G}^{e} =\left(\begin{matrix}
\mathbf{G}^e_{rr} & \mathbf{G}^e_{rs} & \mathbf{G}^e_{rt} \\
\mathbf{G}^e_{sr} & \mathbf{G}^e_{ss} & \mathbf{G}^e_{st} \\
\mathbf{G}^e_{tr} & \mathbf{G}^e_{ts} & \mathbf{G}^e_{tt} 
\end{matrix}\right),    \label{eq:Gdef}
\end{equation}
where each entry, say $\mathbf{G}^{e}_{rr}$, is defined as a diagonal matrix which has entries
\[
(\mathbf{G}^{e}_{rr})_{abc,abc} = w_a w_b w_c G^e_{rr}(r_a,s_b,t_c), 
\]
for $a,b,c = 0,\ldots, N$. We use analogous definitions for the remaining entries of $\mathbf{G}^e$. Using these matrix operators we can write the local stiffness matrix \eqref{eq:stiffQuad} compactly as follows
\[
\mathbf{S}^e = \mathbf{D}^T \mathbf{G}^e \mathbf{D}.
\]
\begin{algorithm*}[t!]
  \caption{BP3.5: collocation differentiation for 3D hexahedral mesh}
  \label{alg:bp35pseudocode}
\begin{boxedminipage}{\textwidth}
    \begin{algorithmic}[1]
    \STATE {{\bf Data:} (1) Vector $\mathbf{q}$, size $N_{el}\times N_p$, (2) differentiation matrix $\mathbf{D}^{\mathrm{1D}}$, size $N_q\times N_q$, (3) geometric factors $\mathbf{G}$, size $N_{el}\times N_p \times 7$, (4)  parameter $\lambda$ };
    \STATE {{\bf Output: } Vector $\mathbf{Sq}$, size $N_{el}\times N_p$ };
   \item[]\COMMENT{Loop over elements $\downarrow$}
    \FOR {$e\in\left\{1,2, \ldots N_{el}\right\}$}
  \FOR{$i,j,k\in\left\{1,2, \ldots N_q\right\}$}
        \item[]\COMMENT{Load geometric factors $\downarrow$}
        \STATE $\texttt{Grr} = \mathbf{G}_{1; kji}^{e}$, $\texttt{Grs} = \mathbf{G}_{2; kji}^{e}$, $\texttt{Grt} = \mathbf{G}_{3; kji}^{e}$; 
        \STATE $\texttt{Gss} = \mathbf{G}_{4; kji}^{e}$, $\texttt{Gst} = \mathbf{G}_{5; kji}^{e}$, $\texttt{Gtt} = \mathbf{G}_{6; kji}^{e}$;
        \item[]\COMMENT{Multiply by $\mathbf{D}$ $\downarrow$}
        \STATE $\texttt{qr} = \sum_{n=1}^{N_q} \mathbf{D}_{in}^{\mathrm{1D}} \mathbf{q}_{kjn}^{e}$;
         \STATE $\texttt{qs} = \sum_{n=1}^{N_q} \mathbf{D}_{jn}^{\mathrm{1D}} \mathbf{q}_{kni}^{e}$;
          \STATE $\texttt{qt} = \sum_{n=1}^{N_q} \mathbf{D}_{kn}^{\mathrm{1D}} \mathbf{q}_{nji}^{e}$;
         \item[] \COMMENT{Apply chain rule $\downarrow$}
		\STATE  $\texttt{rqr}_{ijk}^{e}= \texttt{Grr*qr + Grs*qs + Grt*qt}$;
		\STATE  $\texttt{rqs}_{ijk}^{e}= \texttt{Grs*qr + Gss*qs + Gst*qt}$;
		\STATE  $\texttt{rqt}_{ijk}^{e}= \texttt{Grt*qr + Gst*qs + Gtt*qt}$;
    \ENDFOR
  \FOR{$i,j,k\in\left\{1,2, \ldots N_q\right\}$}
        \STATE $\texttt{J} = \mathbf{G}_{7; kji}^{e}$
    	\STATE  $\mathbf{Sq}_{ijk}^{e} =  \lambda\texttt{J}\mathbf{q}_{kji}^{e}+\sum_{n=1}^{N_q} \mathbf{D}_{ni}^{\mathrm{1D}} \texttt{rqr}_{kjn}^{e} + \mathbf{D}_{nj}^{\mathrm{1D}} \texttt{rqs}_{kin}^{e}+\mathbf{D}_{nk}^{\mathrm{1D}} \texttt{rqr}_{nji}^{e} $;
    \ENDFOR
    \ENDFOR
  \end{algorithmic}
\end{boxedminipage}
\end{algorithm*}
To simplify the action of this local stiffness matrix we again use the fact that the basis interpolation polynomials are a tensor products of one-dimensional polynomials. We define the one-dimensional differentiation matrix $\mathbf{D}^{1D}$ as
\begin{equation*}
\mathbf{D}^{\mathrm{1D}}_{ai} = l_i(r_a), 
\end{equation*}
for all $i,a = 0,\ldots,N$. Using this one-dimensional derivative operator, and the fact that the GLL quadrature nodes collocate with the interpolation nodes using the define the Lagrange basis polynomials $l_i$, we write the partial derivative matrices $\mathbf{D}^r, \mathbf{D}^s,$ and $\mathbf{D}^t$ as tensor products of $\mathbf{D}^{\mathrm{1D}}$ and the identity matrix $I$ as follows
\begin{align*}
\mathbf{D}^r &= \mathbf{D}^{\mathrm{1D}}\otimes I \otimes I, \\
\mathbf{D}^s &= I \otimes\mathbf{D}^{\mathrm{1D}}\otimes I, \\
\mathbf{D}^t &= I \otimes I \otimes\mathbf{D}^{\mathrm{1D}}. 
\end{align*}
Thus, differentiation along each dimension can be computed using single tensor contractions. 

Finally, to write the full action of the local screened Poisson operator $\mathbf{S}^e + \lambda\mathbf{M}^e$ we note that since we have use the collocation GLL quadrature in the evaluation of the integrals \eqref{eq:stiffnessDef}-\eqref{eq:massDef}, we can follow the description of the mass matrix operator above to find that no interpolation is required and the mass matrix can be written simply as $\mathbf{M}^e = \mathbf{J}^e$ where the matrix of geometric factors $\mathbf{J}^e$ is now defined using the GLL weights and quadrature points, i.e.
\[
\mathbf{J}^e_{abc,abc} = w_a w_b w_c |J^e(r_a,s_b,t_c)|,
\]
for $a,b,c = 0,\ldots,N$. Thus we write the local screened Poisson operator as
\[
\mathbf{S}^e + \lambda \mathbf{M}^e = \mathbf{D}^T\mathbf{G}^e\mathbf{D} + \lambda \mathbf{J}^e
\]
This operator can be applied using only six tensor contractions. First, we apply the $\mathbf{D}$ operator by differentiating along each dimension using three tensor contractions. We then multiply by the necessary geometric factors $\mathbf{G}^e$ and apply the transpose operator $\mathbf{D}^T$ with three more tensor contractions. Finally we add the mass matrix contribution which requires no tensor contractions. We detail the full matrix-free action of the screened Poisson operator in the pseudo code in Algorithm \ref{alg:bp35pseudocode}.

\begin{algorithm*}[t!]
  \caption{BP3.0: differentiation for 3D hexahedral elements}
  \label{alg:bp3pseudocode}
\begin{boxedminipage}{\textwidth}
    \begin{algorithmic}[1]
    \STATE {{\bf Data:} (1) Vector $\mathbf{q}$, size $N_{el} \times N_p$, (2) differentiation matrix $\tilde{\mathbf{D}}^{\mathrm{1D}}$, size $N_q^{GL}\times N_q^{GL}$, (3) Interpolation matrix $\mathbf{I}$, size $N_q^{GL}\times N_q$, (4) geometric factors $\mathbf{G}$, size $N_{el} \times N_p^{GL} \times 7$, (5)  parameter $\lambda$ };
    \STATE {{\bf Output: } Vector $\mathbf{Aq}$, size $N_{el} \times N_p$ };
    \FOR {$e\in\left\{1,2, \ldots N_{el}\right\}$}
    \STATE $\tilde{\mathbf{q}}^{e}= \mathrm{Interpolate}(\mathbf{q}^e,\mathbf{I}^{\mathrm{1D}})$ \COMMENT{Interpolate to GL nodes (Algorithm \ref{alg:interp})}\;
    \FOR{$i,j,k\in\left\{1,2, \ldots N_q^{GL}\right\}$}
         \item[]\COMMENT{Load geometric factors $\downarrow$}
        \STATE $\texttt{Grr} = \mathbf{G}_{1; kji}^{e}$, $\texttt{Grs} = \mathbf{G}_{2; kji}^{e}$, $\texttt{Grt} = \mathbf{G}_{3; kji}^{e}$ ;
        \STATE $\texttt{Gss} = \mathbf{G}_{4; kji}^{e}$, $\texttt{Gst} = \mathbf{G}_{5; kji}^{e}$, $\texttt{Gtt} = \mathbf{G}_{6; kji}^{e}$;
         \item[]\COMMENT{Multiply by $\tilde{\mathbf{D}}$ $\downarrow$}
        \STATE $\texttt{qr} = \sum_{n=1}^{N_q^{GL}} \tilde{\mathbf{D}}^{\mathrm{1D}}_{in}   \tilde{\mathbf{q}}_{kjn}^{e}$;
         \STATE $\texttt{qs} = \sum_{n=1}^{N_q^{GL}} \tilde{\mathbf{D}}^{\mathrm{1D}}_{jn}  \tilde{\mathbf{q}}_{kni}^{e}$;
          \STATE $\texttt{qt} = \sum_{n=1}^{N_q^{GL}} \tilde{\mathbf{D}}^{\mathrm{1D}}_{kn} \tilde{\mathbf{q}}_{nji}^{e}$;
              \item[] \COMMENT{Apply chain rule $\downarrow$}
		\STATE  $\texttt{rqr}_{ijk}^{e}= \texttt{Grr*qr + Grs*qs + Grt*qt}$;
		\STATE  $\texttt{rqs}_{ijk}^{e}= \texttt{Grs*qr + Gss*qs + Gst*qt}$;
		\STATE  $\texttt{rqt}_{ijk}^{e}= \texttt{Grt*qr + Gst*qs + Gtt*qt}$;
    \ENDFOR
          \FOR{$i,j,k\in\left\{1,2, \ldots N_q^{GL}\right\}$}
        \STATE $\texttt{J} = \mathbf{G}_{7; kji}^{e}$
    	\STATE  $\tilde{\mathbf{Aq}}_{ijk}^{e} =  \lambda\texttt{J}\mathbf{q}_{kji}^{(e, GL)}+\sum_{n=1}^{N_q^{GL}} \tilde{\mathbf{D}}^{\mathrm{1D}}_{ni} \texttt{rqr}_{kjn}^{e} + \tilde{\mathbf{D}}^{\mathrm{1D}}_{nj} \texttt{rqs}_{kin}^{e}+\tilde{\mathbf{D}}^{\mathrm{1D}}_{nk} \texttt{rqr}_{nji}^{e} $;
    \ENDFOR
    \STATE $\mathbf{Aq}^{e}= \mathrm{Project}(\tilde{\mathbf{Aq}}^e,\mathbf{I}^{\mathrm{1D}})$ \COMMENT{Project to GLL nodes (Algorithm \ref{alg:proj})}\;
    \ENDFOR
  \end{algorithmic}
\end{boxedminipage}
\end{algorithm*}

\subsection{BP3.0: Stiffness matrix evaluated with quadrature}

The final benchmark we consider is the same matrix-vector product of the high-order screened Poisson operator $\mathbf{S}+\lambda\mathbf{M}$ as in BP3.5. This time, however, we use the full GL quadrature to approximate the integrals in \eqref{eq:stiffnessDef} and \eqref{eq:massDef}. This benchmark combines computational elements from BP1.0 and BP3.5 which makes for a more arithmetically intense kernel and maximizing its performance on GPUs is challenging.

We again begin by describing the local stiffness matrix $\mathbf{S}^e$ defined in \eqref{eq:stiffnessDef}. We evaluate the integral in \eqref{eq:stiffnessDef} in each dimension separately, this time using the full GL interpolation nodes as the quadrature. Using the notation introduced in BP1.0 above, we write
\begin{equation}
\mathbf{S}^e_{ijk,i'j'k'} = \sum_{a=1}^{N_q^{GL}} \sum_{b=1}^{N_q^{GL}} \sum_{c=1}^{N_q^{GL}} \tilde{w}_a \tilde{w}_b \tilde{w}_c (\nabla l_{i'j'k'}(\tilde{r}_a,\tilde{s}_b,\tilde{t}_c))^T  G^{e}(\tilde{r}_a,\tilde{s}_b,\tilde{t}_c) \nabla l_{ijk}(\tilde{r}_a,\tilde{s}_b,\tilde{t}_c). \label{eq:stiffQuad2}
\end{equation}
Note here that the quadrature requires the gradients of the basis polynomials $l_{ijk}$ to be evaluated at the GL quadrature nodes. Were we to simply compose the interpolation and differentiation operators $\mathbf{I}$ and $\mathbf{D}$ defined in BP1.0 and BP3.5 above we would require nine tensor contractions to evaluate this quantity. Indeed, for each of the $r$, $s$, and $t$ derivatives of $l_{ijk}$, we would require an operation which combines differentiation and interpolation to the GL quadrature along one dimension and only interpolation to the GL quadrature along the remaining two dimensions. 

We instead reduce the number of required tensor contractions by considering the Lagrange interpolating polynomials $\tilde{l}_a(r)$ for $a = 1,\ldots, N_q^{GL}$ which interpolate the GL quadrature nodes. We define the tensor product basis polynomials as done for the GLL interpolating basis as
\[
\tilde{l}_{abc}(r,s,t) = \tilde{l}_a(r)\tilde{l}_b(s)\tilde{l}_c(t),
\]
for $a,b,c = 1,\ldots,N_q^{GL}$. We then define the derivative operators, as done above for the GLL interpolating Lagrange basis functions, on this set of polynomials as follows 
\begin{align*}
\tilde{\mathbf{D}}^r_{abc,a'b'c'} &= \frac{\partial\tilde{l}_{a'b'c'}}{\partial r}(\tilde{r}_a,\tilde{s}_b,\tilde{t}_c), \\
\tilde{\mathbf{D}}^s_{abc,a'b'c'} &= \frac{\partial\tilde{l}_{a'b'c'}}{\partial s}(\tilde{r}_a,\tilde{s}_b,\tilde{t}_c), \\
\tilde{\mathbf{D}}^t_{abc,a'b'c'} &= \frac{\partial\tilde{l}_{a'b'c'}}{\partial t}(\tilde{r}_a,\tilde{s}_b,\tilde{t}_c),
\end{align*}
for $a,b,c,a',b',c' = 1,\ldots,N_q$. We can then construct the gradient operation on this set of basis function as $\tilde{\mathbf{D}} = [\tilde{\mathbf{D}}^r,\tilde{\mathbf{D}}^s,\tilde{\mathbf{D}}^t]^T$. Thus, if we view the interpolation operators defined in \eqref{eq:interpolationOps} as transforming a polynomial from the basis of Lagrange interpolating polynomials on the GLL nodes to the basis of Lagrange interpolating polynomials on the GL quadrature then we can use the operator $\tilde{\mathbf{D}}$ to evaluate the derivatives of this polynomial on the GL node basis. 

With these derivative operators, we continue as in BP3.5 by defining the matrix of weights and geometric data $\mathbf{G}^e$ in \eqref{eq:Gdef} where this time 
\[
(\mathbf{G}^{e}_{rr})_{abc,abc} = \tilde{w}_a \tilde{w}_b \tilde{w}_c G^e_{rr}(\tilde{r}_a,\tilde{s}_b,\tilde{t}_c), 
\]
for $a,b,c =1,\ldots,N_q^{GL}$. We use analogous definitions for the remaining entries of $\mathbf{G}^e$. Using these matrix operators we can write the local stiffness matrix \eqref{eq:stiffQuad} compactly as follows
\[
\mathbf{S}^e = \mathbf{I}^T\tilde{\mathbf{D}}^T \mathbf{G}^e \tilde{\mathbf{D}}\mathbf{I}.
\]
Note that, as with the GLL interpolation basis polynomials, we can define the one-dimensional differentiation operator $\tilde{\mathbf{D}}^{\mathrm{1D}}$ defined such that 
\[
\tilde{\mathbf{D}}^{\mathrm{1D}}_{ai} = \tilde{l}_i(\tilde{r}_a), 
\]
for $i,a = 1,\ldots N_q^{GL}$, so that each differentiation operation in $\tilde{\mathbf{D}}$ can be written as a tensor product of $\tilde{\mathbf{D}}^{\mathrm{1D}}$ and the identity matrix $I$. Therefore the differentiation operators on the GL Lagrange interpolation basis polynomials along each dimension can be applied using a single tensor contraction. 

Finally, to express the full action of the local screened Poisson operator $\mathbf{S}^e + \lambda\mathbf{M}^e$ we combine the discussion in BP1.0 regarding using the GL quadrature to evaluate the local mass matrix $\mathbf{M}^e$ in order to write the local screened Poisson operator as
\[
\mathbf{S}^e + \lambda \mathbf{M}^e =  \mathbf{I}^T\tilde{\mathbf{D}}^T \mathbf{G}^e \tilde{\mathbf{D}}\mathbf{I} + \lambda \mathbf{I}^T \mathbf{J}^e \mathbf{I},
\]
where $\mathbf{J}^e$ is defined as in BP1.0 above. This operator can be applied using a total of twelve tensor contractions. We first interpolate to the GL quadrature nodes by applying the interpolation operator $\mathbf{I}$ using three tensor contractions. We then differentiate along each dimension by applying the $\tilde{\mathbf{D}}$ operator using three tensor contractions. We then multiply by the necessary geometric factors and multiply by the transpose derivative operator using three additional tensor contractions and add the mass matrix contributions. Finally we use three tensor contractions to multiply by the transpose interpolation operator in each dimension to project the result back to the GLL interpolation nodes. We detail the full matrix-free action of the stiffness matrix evaluated with the numerical quadrature in pseudo code in Algorithm \ref{alg:bp3pseudocode}.

\section{Empirical performance model} \label{performance.sec}

The performance for all three benchmark problems is limited by global memory bandwidth. In BP3.5 and BP3.0 we load seven geometric factors for every node in the FEM element in addition to loading and storing the field vector $\mathbf{q}$ itself. Moreover, as we emphasize in the introduction, for all three operations the data-to-FLOP ratio is high which limits our ability to hide memory latency behind computation.

\citep{Konstantinidis2015} proposed a GPU performance model that accounts for the various data-to-flop ratios. However, the kernels used in the model achieve occupancy of around $97\%$ while, for our kernels, extensive use of register file and shared memory results in achieved occupancy rarely exceeding 25\%. Hence, we cannot apply this model for our kernels in a meaningful way. 

A different idea was presented in \citep{Abdelfattah2016}, where the authors used the size of global memory transfers as a means of comparison. The advantage of such approach is that it is independent of implementation. In this work, we use a model which is similar to \citep{Abdelfattah2016}. However, we transfer comparatively more data due to the geometric factors. Our model is also based on an assumption that the bandwidth of the global device data transfer is the limiting factor. Thus, we compare the performance of our kernels to the performance of copying data from global GPU memory to global GPU memory. The size of the data transfer we compare to is equivalent to the size of the data moved from and to the global memory in the particular benchmark problems, see Table~\ref{tab:tab1}.

For example, for every element in the FEM mesh the BP1.0 code needs to read 
\[
R = N_p + N_p^{GL}
\]
doubles and needs to write 
\[
W = N_p
\]
doubles. Hence, the total global memory transfer is 
\[
T_{\mathrm{BP1.0}} = 2N_p + N_p^{GL},
\] 
bytes of data per element. Since the memory bus is bi-directional and a double variable consists of eight bytes of data, we compare the performance of our code for BP1.0 to transferring 
\[
8 N_{el}\frac{2N_p + N_p^{GL}}{2}
\]
bytes of data. Each data transfer is executed ten times using a standard \verb|cudaMemcpy| function and the performance in terms of GB/s is measured by taking an average of the ten measurements.

\begin{table}
\begin{center}
\def\arraystretch{1}
{\small
\begin{tabular}{lccc}
\hline
 &BP1 & BP3.5& BP3 \\ 
 \hline
  Read (R)  & $N_p + N_p^{GL}$ & $8N_p$ & $N_p + 7N_p^{GL}$\\
  Write (W) & $N_p$ &$N_p$& $N_p$\\
  Total bytes (T) & $2N_p+N_p^{GL}$ & $9N_p$  &$2N_p + 7N_p^{GL}$\\
  \hline
\end{tabular}}
\end{center}
\caption{The minimum number of doubles read and written to the global GPU memory per element in the three problems considered in this paper. \label{tab:tab1}}
\end{table} 

The efficiency of data transfer depends on the size of data, with  throughput maximized if sufficiently large amounts of data are being transferred. We therefore expect higher bandwidth for a mesh with more elements, and for higher degree polynomial approximations. We note that for the GPU used in this paper, the NVIDIA Tesla P100 12 GB PCI-E, the mesh containing 512 elements is too small to effectively hide data transfer latencies. To see this, we note that our approach is to parallelize the problem by assigning each element to a thread block. The NVIDIA Tesla P100 has 56 SMs and two blocks of threads can be processed simultaneously on every SM. The GPU is therefore capable of processing 112 blocks of threads concurrently, which for the mesh of 512 elements means only 5 thread blocks are executed per SM. The result of this small work load per SM is that the overhead cost of kernel launch is not offset by the execution time of the kernel itself and also the empirical bandwidths which we observe in these data transfers are noisy due to overhead costs. 

Let $d_r$ denote the number of bytes read from the global GPU memory and  $d_w$ denote the number of bytes written to global GPU memory by a GPU kernel. Let $B_{gl}$ denote global memory bandwidth for copying $\frac{d_r+d_w}{2}$ bytes of data. Let $F$ denote the number of floating point operations that must be executed in this kernel. In our model, the maximal GFLOPS/s are estimated using a formula
\begin{equation} \label{eq:globalRoof}
\mathcal{R}_{\text{global}} = \frac{B_{gl} \cdot F}{d_w+d_r}.
\end{equation}

\begin{figure}[t]
\centering
\includegraphics[width=.4\textwidth]{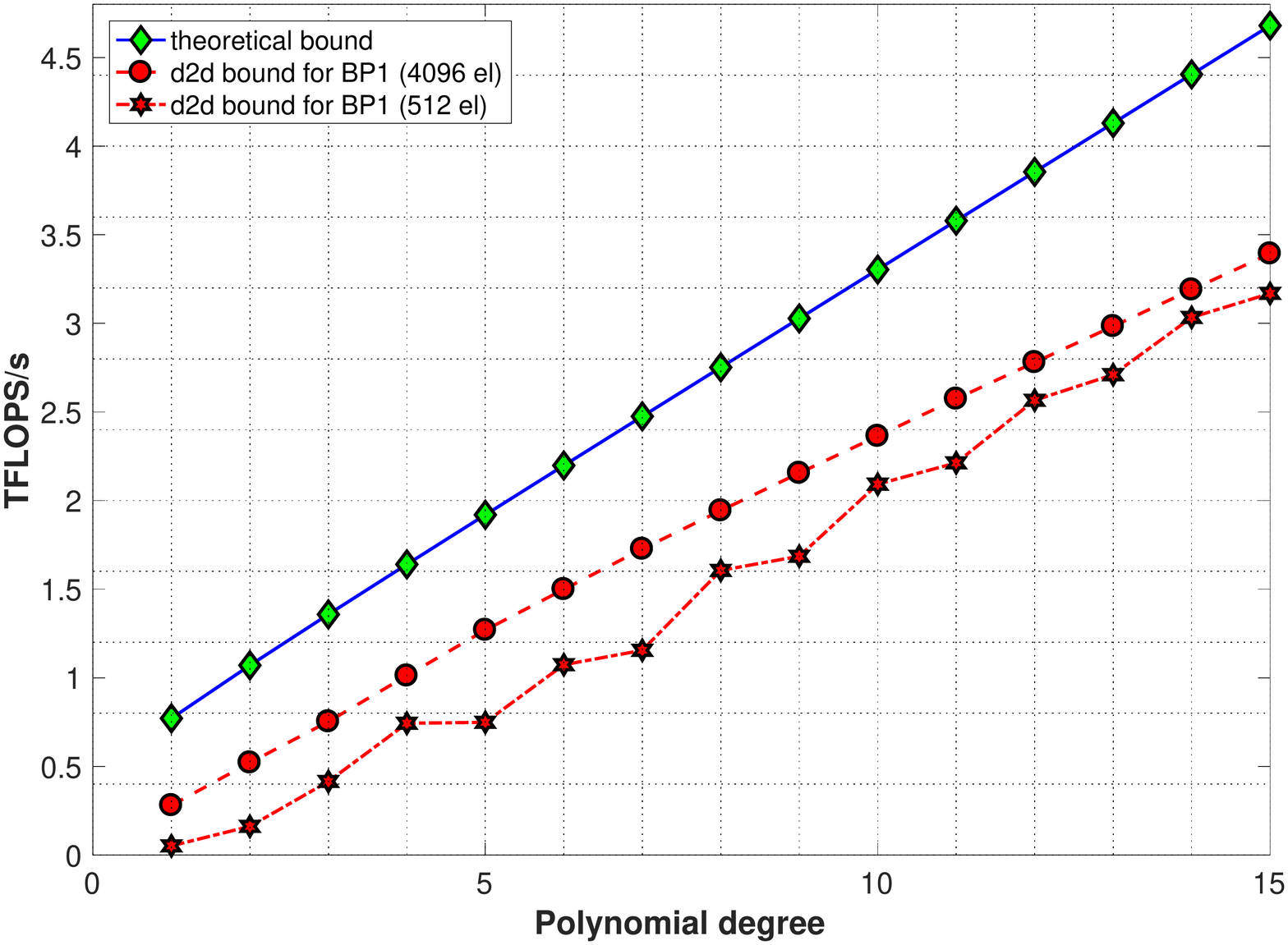}\includegraphics[width=.4\textwidth]{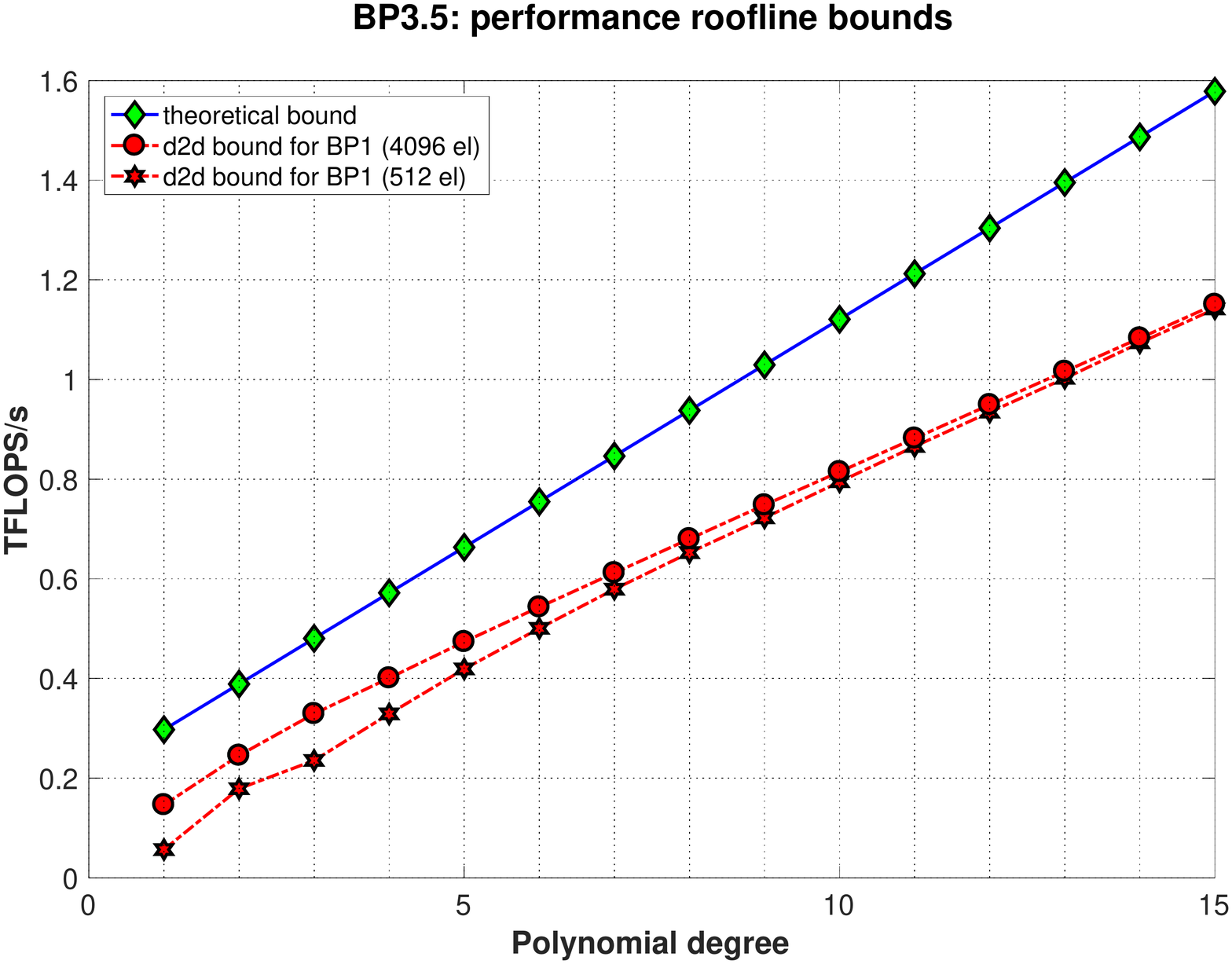} 

\includegraphics[width=.4\textwidth]{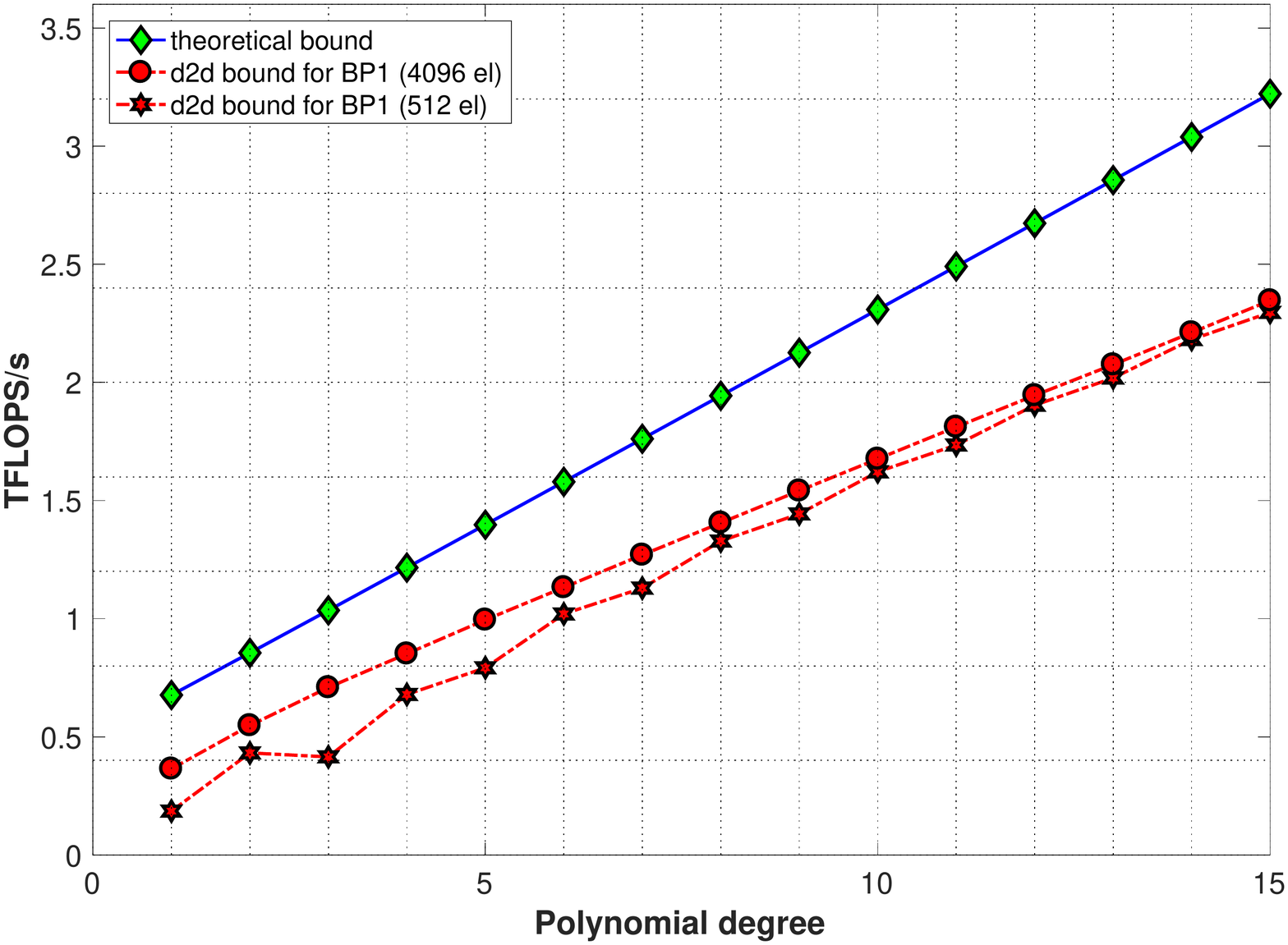}
\caption{Performance roofline bounds for the BP1.0 (top left), BP3.5 (top right), and BP3.0 (bottom) benchmarks. In each chart the upper plot (line with diamond-shaped ticks) shows a theoretical bound obtained using theoretical peak bandwidth of $549$ GB/s for the NVIDIA P100 PCI-E 12GB GPU. The lower plots show the empirical peak bandwidth bound obtained using measured bandwidth attained when performing a device memory to device memory copy (upper line for a hexahedral mesh with 4096 elements and lower line for a hexahedral mesh with 512 elements).}
\label{fig:boundsBP}
\end{figure}

For BP1.0, BP3.5 and BP3.0 the roofline $\mathcal{R}_{\text{global}}$ is evaluated as a function of polynomial degree $N$. Figure~\ref{fig:boundsBP} shows maximum TFLOPS/s for BP1.0, BP3.5 and BP3.0, respectively. 

Inspired by~\citep{Volkov2008},  for several kernels tested for BP1.0 and BP3.0., we devised a supplementary theoretical roofline based on shared memory bandwidth. We observe that in addition to copying the data to/from global memory, we also copy the data to/from shared memory\footnote{Unlike the copy-based empirical streaming roofline, the shared memory performance  bound depends on the kernel, not on the problem}. Since the memory bandwidth of shared memory is lower than the register bandwidth, shared memory transactions can limit performance. We denote the shared memory bandwidth by $B_{sh}$ and and estimate it with 
\begin{equation*}
\begin{split}
B_{\text{sh}} &= \text{\#SMs} \times \text{SIMD width} \\
&\times \text{word length}  \times \text{clock speed in Ghz}.
\end{split}
\end{equation*}
With this ansatz we estimate that the shared memory for the NVIDIA Tesla P100 12 GB PCI-E GPU is limited to  $B_{sh} = 56\cdot 32 \cdot 4 \text{ bytes}  \cdot 1.328 \text{Ghz} = 9.5191 \text{TB/s}$. The shared memory roofline model is then given by the equation
\begin{equation}\label{eq:sharedRoof}
\mathcal{R}_{\text{shared}} = \frac{B_{\text{sh}} \cdot F}{s_r +s_w},
\end{equation}
where $F$ denotes the number of floating point operations performed (per thread block), $s_r$ is the number of bytes read from the shared memory (per thread block) and $s_w$ is the number of bytes written to the shared memory (per thread block).



\section{Optimizing Benchmark Problem 1.0} \label{BP1.sec}
In this section we describe the sequence of optimization strategies that were applied to the implementation of the BP1 operation.

\begin{figure*}
    \centering
    \includegraphics[scale =0.2]{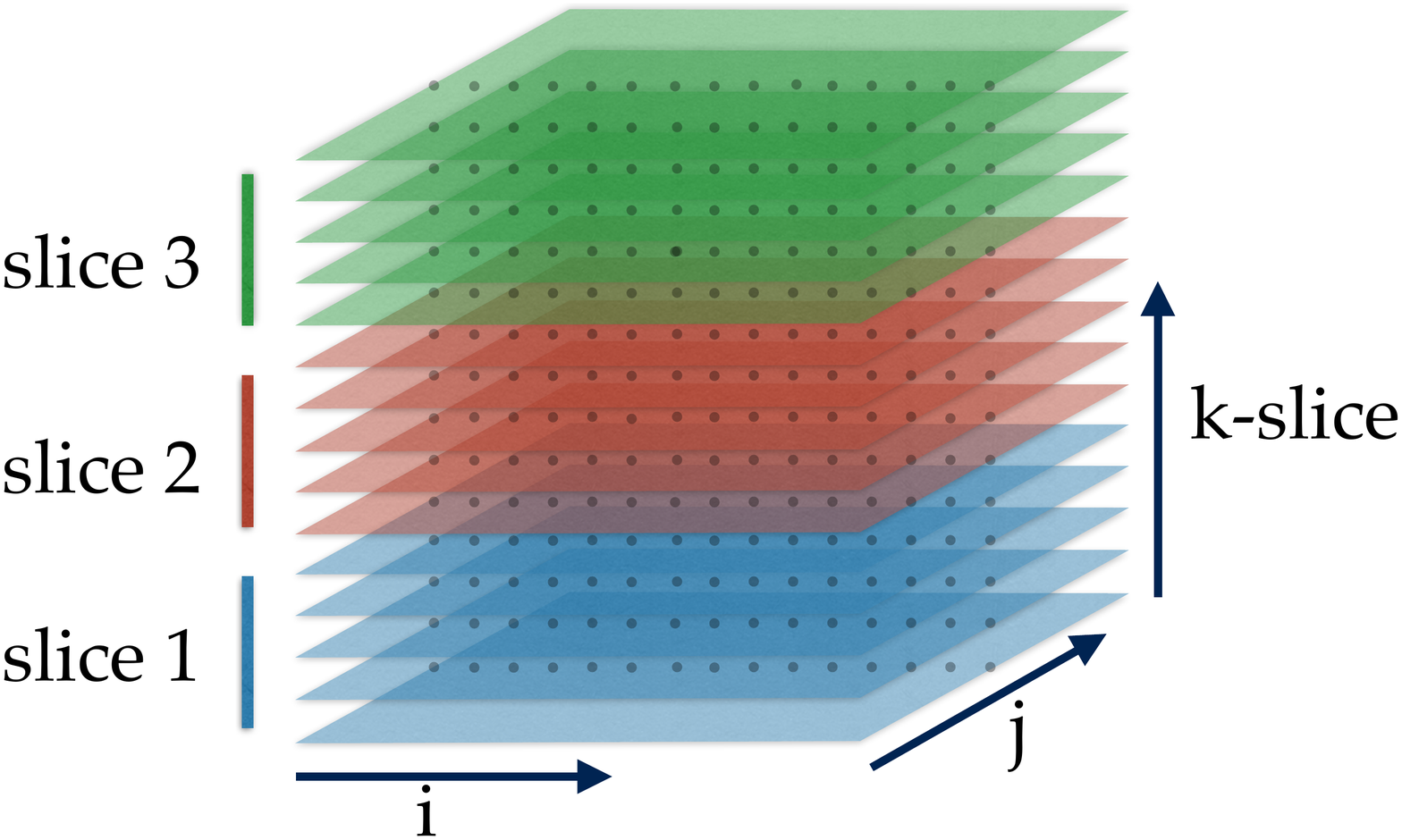}
    \includegraphics[scale=0.15]{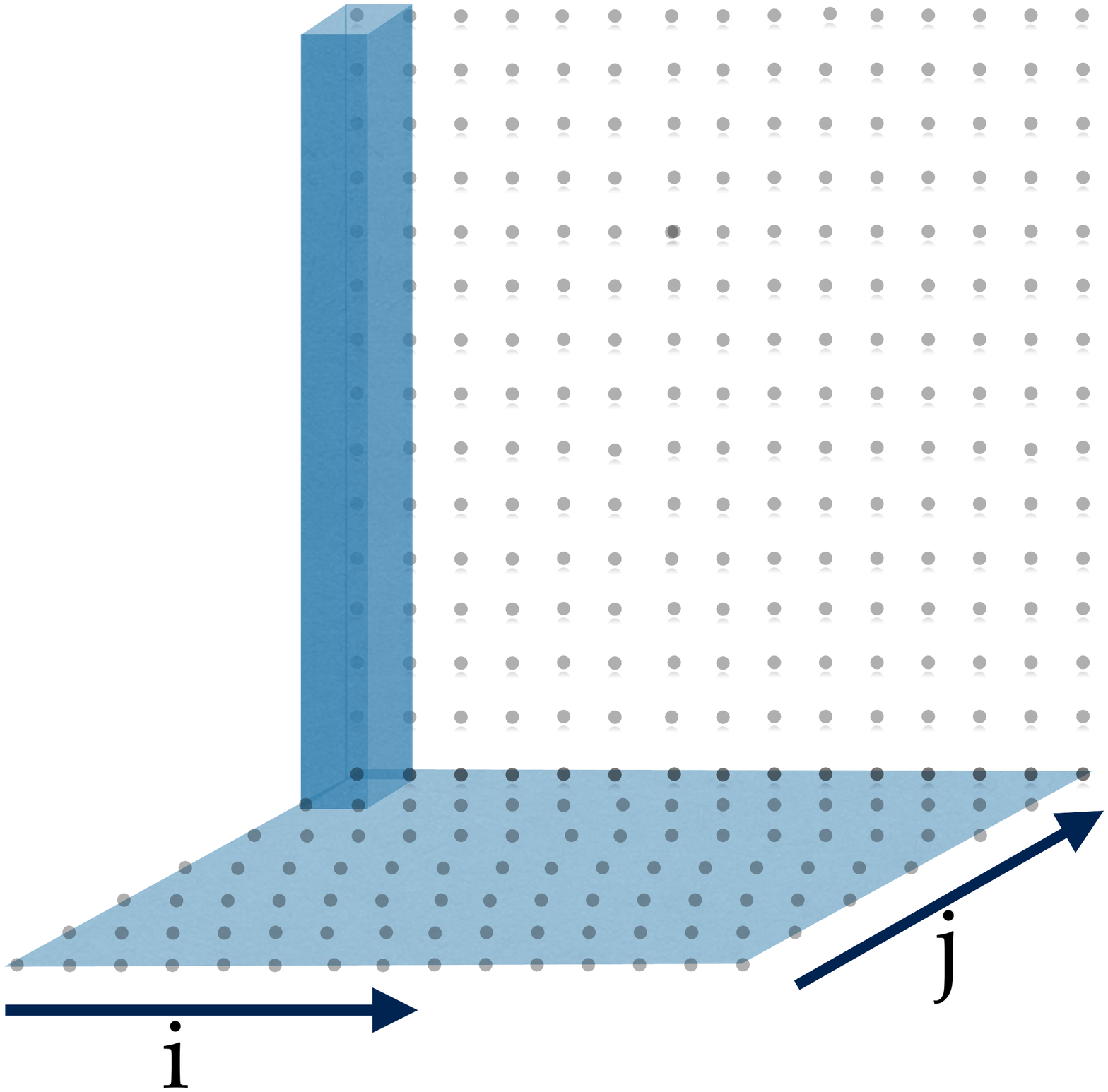}
    \caption{3D vs 2D thread structure. On the left: 3D approach -- each thread processes a ``slice'' of nodes. On the right: 2D approach -- each thread processes a vertical ``column'' of nodes. }
    \label{fig:2Dvs3D}
\end{figure*}

\noindent {\bf Kernel design.} Recall that the tensor contraction operations for each element in the FEM mesh can be performed independently of other elements. Thus, to parallelize the FEM operations we assign each element to a block of threads on the GPU. In a previous work, \citep{Remacle2016} associated a single node of an element with a single thread. This is, however, impossible for high-order interpolation because if $N_q\geq 9$, we exceed the maximum number of threads per block of threads (currently limited by CUDA to $1024$). Hence, for higher-order approximations we need to assign multiple nodes to a thread. We can subdivide the nodes in each element using either use a 3D thread structure or a 2D thread structure. Figure~\ref{fig:2Dvs3D} shows two such approaches. For BP1.0, we use a 2D thread structure since we found it more effective. We also investigate a 3D thread structure for $N_q\leq 9$ for BP3.5 and $N_q<9$ for BP3.0 

In BP1.0 the action of the mass matrix $\mathbf{M}$ on each element can be applied using six tensor constraction operations. Hence, Algorithm~\ref{alg:bp1alg} consists of $6$ contractions wherein we cycle through the entire $\mathbf{q}^{e}$. Block-wise synchronization is needed between the contractions to ensure that the previous operation has completed. At the minimum, we need to enforce this block synchronize five times. Using a 2D thread structure requires more thread synchronizations because we process only a slice of the nodes at a time.

\noindent{\bf BP1.0 thread memory optimization.} Because the one-dimensional interpolation matrix $\mathbf{I}^{\mathrm{1D}}$ is used by all the threads in the block, we load $\mathbf{I}^{\mathrm{1D}}$ into the shared memory. For the field variable $\mathbf{q}$, we can either load $\mathbf{q}^{e}$ to shared memory, fetch it to registers, or fetch it piece-by-piece from global memory when needed. Note that we also need a placeholder array to store the partial results between the loops. There are several options to choose from, however only a single array of size $N^{GL}_p$ can be stored in shared memory. Storing two such arrays is not feasible since we exceed the limit of 48 KB shared memory for thread block for a large $N$. 

\begin{figure*}[t]
\centering
\includegraphics[width=0.45\textwidth]{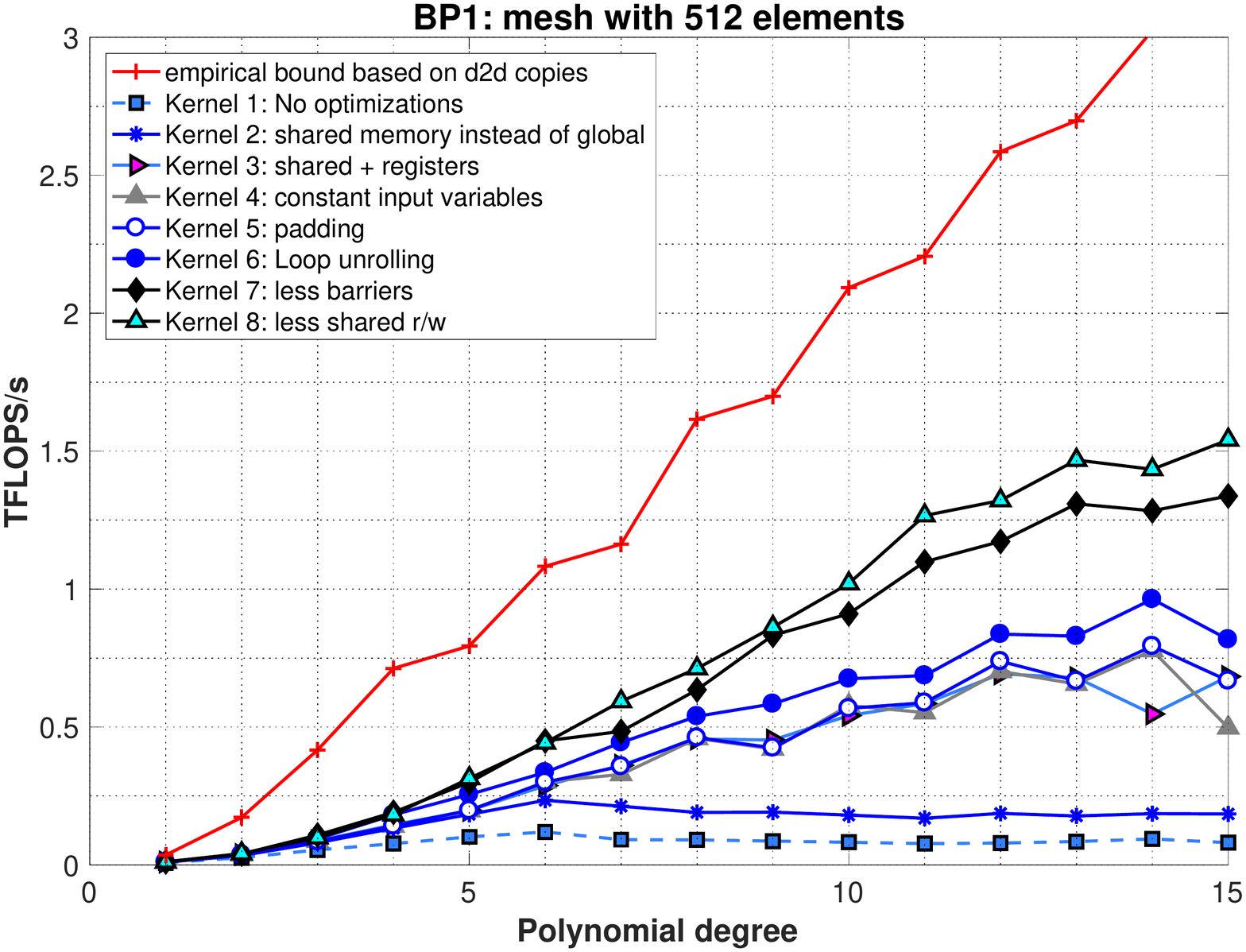} \includegraphics[width=0.45\textwidth]{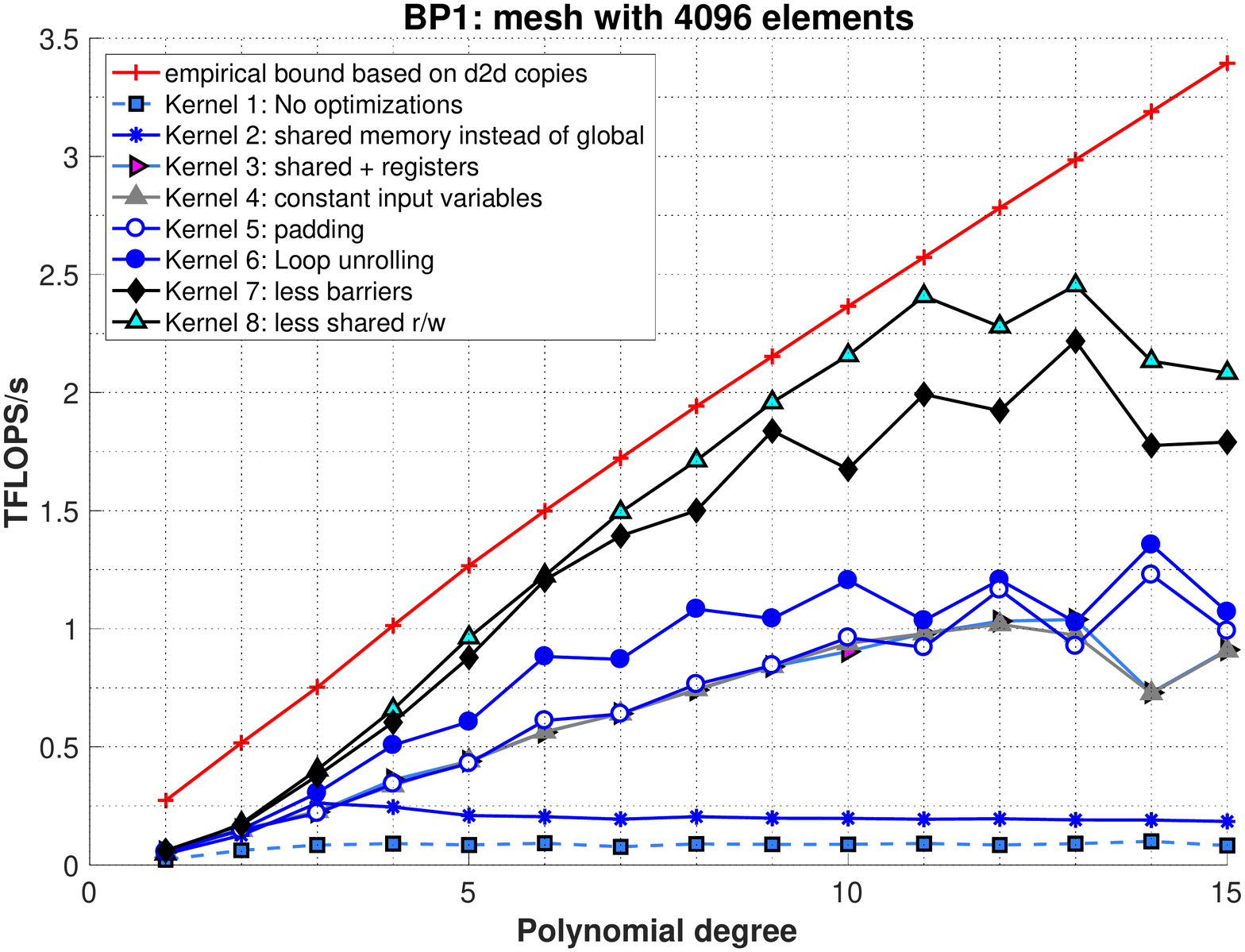}
\caption{BP1.0: achieved floating point performance. Left: results obtained using cube-shaped mesh with 512 elements. Right: results obtained using cube-shaped mesh with 4096 elements on a NVIDIA P100 PCI-E 12GB GPU. }
\label{fig:bp1res}
\end{figure*}

\noindent{\bf BP1.0 kernel optimization.} We show the performance results of eight GPU kernels in Figure~\ref{fig:bp1res}. The eight kernels were constructed in sequence beginning from a direct implementation of the pseudo-code in Algorithm \ref{alg:bp1alg} and applying successive optimizations. The results shown in Figure~\ref{fig:bp1res} help to demonstrate the effect each optimization has on the overall performance of BP1.0. We list below some details of each kernel here as well as any optimization made in that kernel. Unless otherwise noted, Kernel $n$ contains all the optimizations contained in Kernel $n-1$. 

\begin{figure}[t]
\centering
\includegraphics[width=0.4\textwidth]{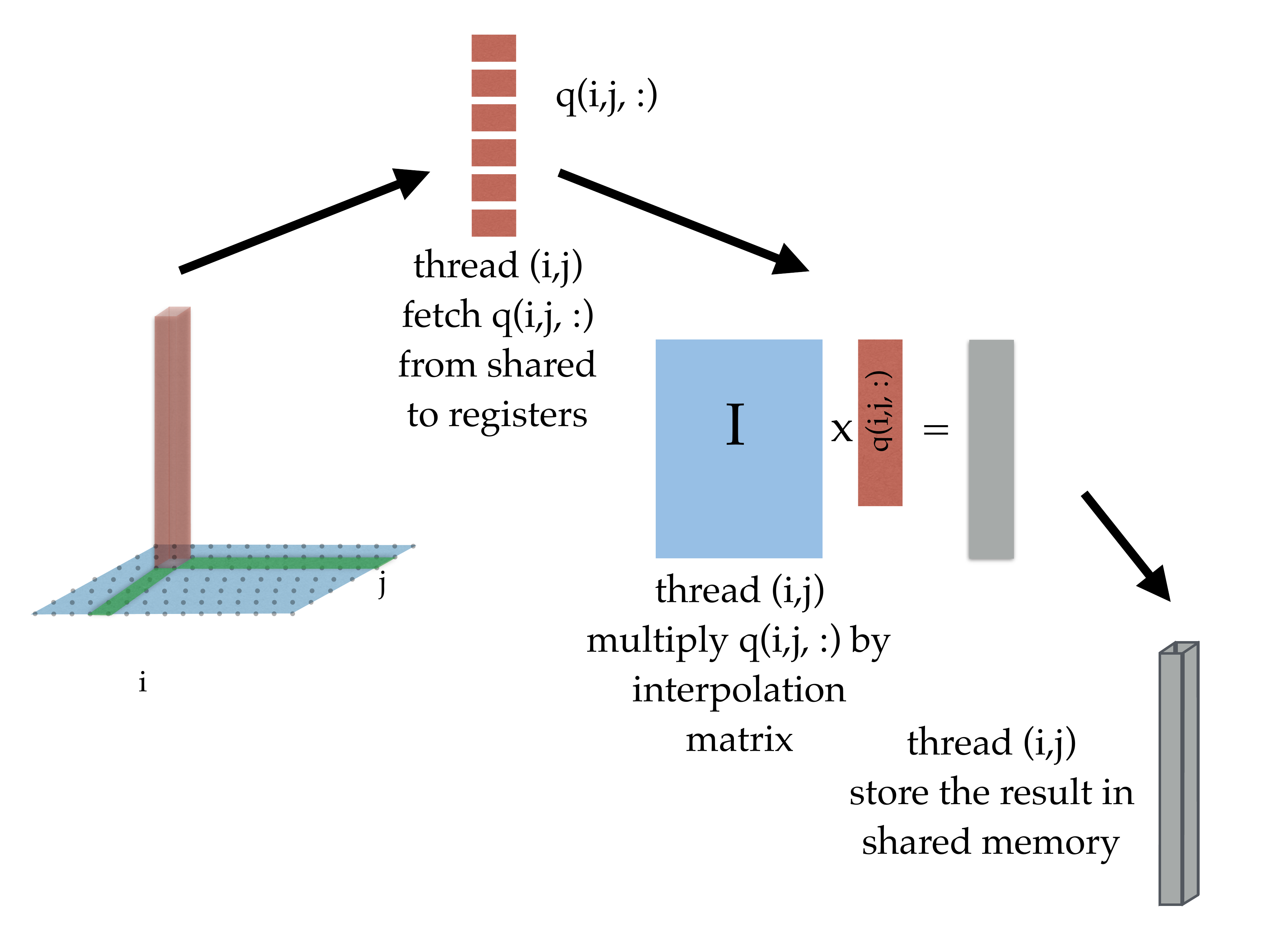}
\caption{BP1.0: the idea behind reducing synchronizations in Kernel 7. We fetch pieces of $\mathbf{q}^{e}$ to registers from shared memory and then write the result to shared memory. This does not create race conditions because we do use a 2D thread structure and interpolate only in one direction at a time. }
\label{fig:bp1v2idea}
\end{figure}

\kernelfl{1} This kernel serves as a reference implementation. It uses a 2D thread structure associated with horizontal $(i,j)$ slices (see right side of Figure \ref{fig:2Dvs3D}). We declare two additional global memory variables for storing intermediate results. Kernel 1 only uses shared memory for the interpolation matrix. In all the loops, it reads from and writes to global memory. As a result, the reference kernel is highly inefficient, reaching only 80 GFLOPS/s even for the larger mesh. 
\kernel{2} In this kernel the global auxiliary arrays are replaced by two shared memory arrays  (with $\left(N^{GL}_q\right)^2$ elements in each array). While processing $\mathbf{q}^{e}$, we read directly from the input arrays, without caching to shared memory or registers. Due to the reduced number of global memory fetches, the performance  improves by approximately a factor of two.
\kernel{3} In this kernel each thread allocates a register array of size $N_q^{(GL)}$ and copies a section of $\mathbf{q}^{e}$ to the array at the beginning of the kernel. This kernel reaches 1 TFLOP/s in the best case. 
\kernel{4} In this kernel all the input variables, except the variable to which the output is saved, are declared as \verb|const|. This yields only a marginal improvement in performance.
\kernel{5} For this kernel, if $N_q =8$, or $16$ or  $N_q^{GL} =8$, or $16$, we pad the shared memory arrays used for storing $\mathbf{I}^{\mathrm{1D}}$ and partial results to avoid bank conflicts. There is only a noticeable improvement for $N=15$ for the smaller mesh and $N=14$ for the larger mesh. 
\kernel{6} In this kernel all loops, including the main loop in which we process the $(i,j)$ slices are unrolled. Unrolling the $(i,j)$ loop is important for the performance, since it gives the scheduler more freedom and more opportunity for instruction-level parallelism. At this point, the performance exceeds one TFLOP/s for the small mesh and 1.4 TFLOPS/s for the large mesh. 
\kernel{7} All kernels presented thus far have required $5N_q^{GL}+1$ thread synchronizations in a block. For example with $N=12$, for which $N_q^{GL} = 14$, the number of synchronizations is $71$. In this kernel the number of thread synchronizations is reduced to five by allocate more shared memory. That is, we load the entire $\mathbf{q}^{e}$ array to shared memory as opposed to only loading $\mathbf{q}^{e}$ slice-by-slice. Figure~\ref{fig:bp1v2idea} illustrates the idea behind reducing synchronizations. Kernel 7 brings the performance up to 2.25 TFLOPS/s,
\kernel{8}: \citep{volkov-2010} emphasized that shared memory is much slower than using registers. Hence, reducing total number of read and write requests to/from shared memory and replacing them with register read/write instructions can significantly improve performance. Kernel 8 exploits the structure of the matrix $\mathbf{I}^{\mathrm{1D}}$ to reduce the number of shared memory transactions. Specifically, each entry in $\mathbf{I}^{\mathbf{1D}}$ appears twice: $\mathbf{I}^{\mathbf{1D}}_{nm} = \mathbf{I}^{\mathbf{1D}}_{N_q^{GL}-n+1, N_q-m+1}$. 
Exploiting this symmetry lets us halve the number of loads from $\mathbf{I}^{\mathbf{1D}}$. This kernel loads the entire $\mathbf{I}^{\mathbf{1D}}$ into shared memory and inside each loop an appropriate entry is copied from shared memory to a register once and used twice. Since in BP1.0 we need to multiply by $\mathbf{I}^{\mathbf{1D}}$ and by $(\mathbf{I}^{\mathbf{1D}})^T$, pre-fetching only a half of matrix $\mathbf{I}^{\mathbf{1D}}$ would complicate the code and require a set of additional if-statements, which cause thread divergence. The resulting reduction in shared memory operations brings the measured performance close to the empirical roofline.

\noindent{\bf BP1.0 results.} The performance numbers presented in Figure~\ref{fig:bp1res} reveal that the kernels considered can be categorized into three groups. Kernels 1 and 2 form the first group, Kernels 3--6 form the second group and Kernels 7 and 8 form the third group. Between each of these three groups we observe significant performance improvements. The first major improvement occurs for Kernel 3 and is the result of cashing $\mathbf{q}^{e}$ to register arrays prior to contraction. The second significant improvement occurs for Kernel 7 and is due to reducing the number of barriers. The performance of our reference kernel, Kernel 1, in the best case reaches only 80 GFLOPS/s, whereas our best kernel, Kernel 8, achieves 2.5 TFLOPS/s, yielding a 31 fold speedup. To obtain more than two TFLOPS/s, we needed to change the algorithm to account for the limitations intrinsic to the GPU, namely the cost of many thread synchronizations.

\begin{figure}[t]
\centering
\includegraphics[width=.4\textwidth]{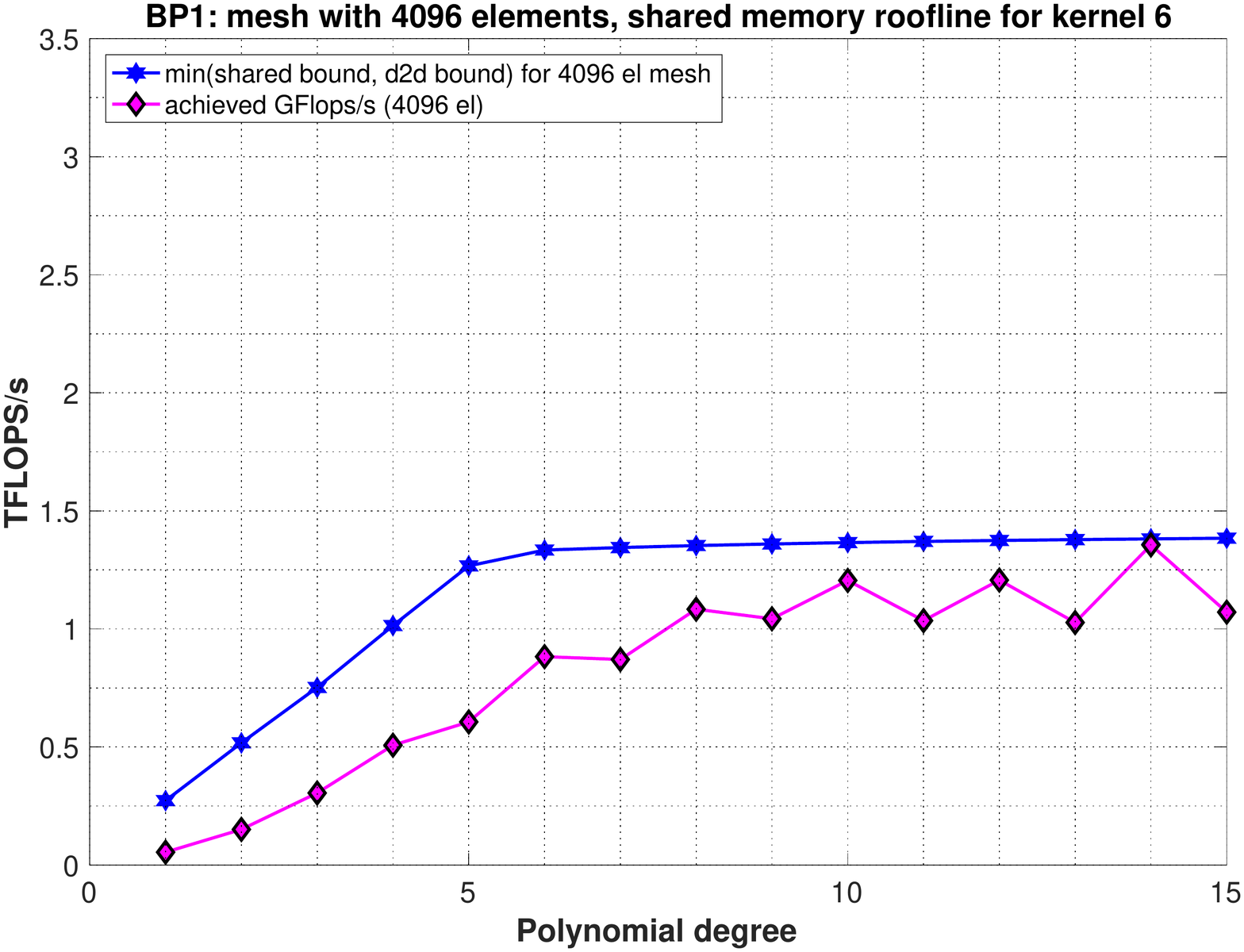}
\includegraphics[width=.4\textwidth]{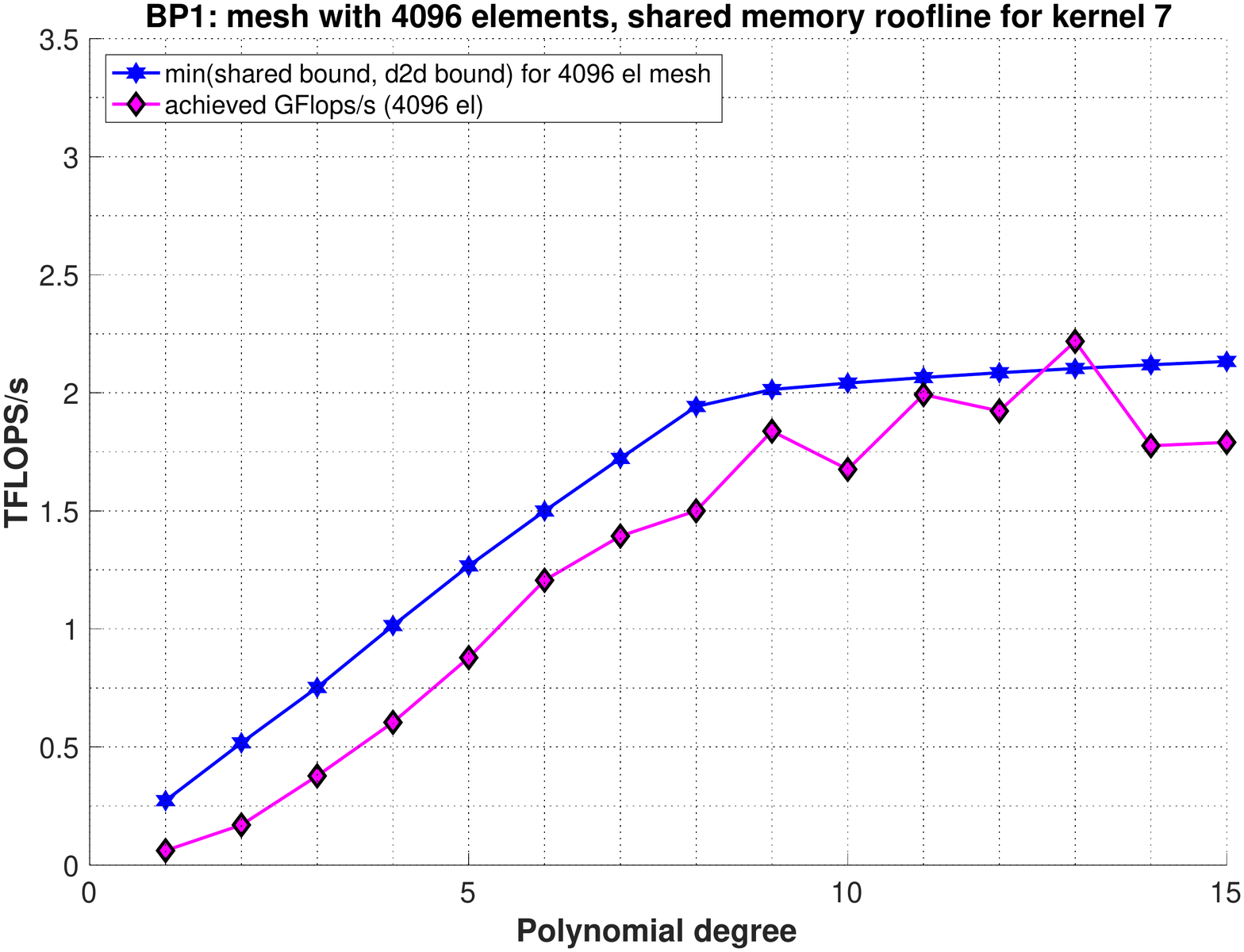}\\
\includegraphics[width=.4\textwidth]{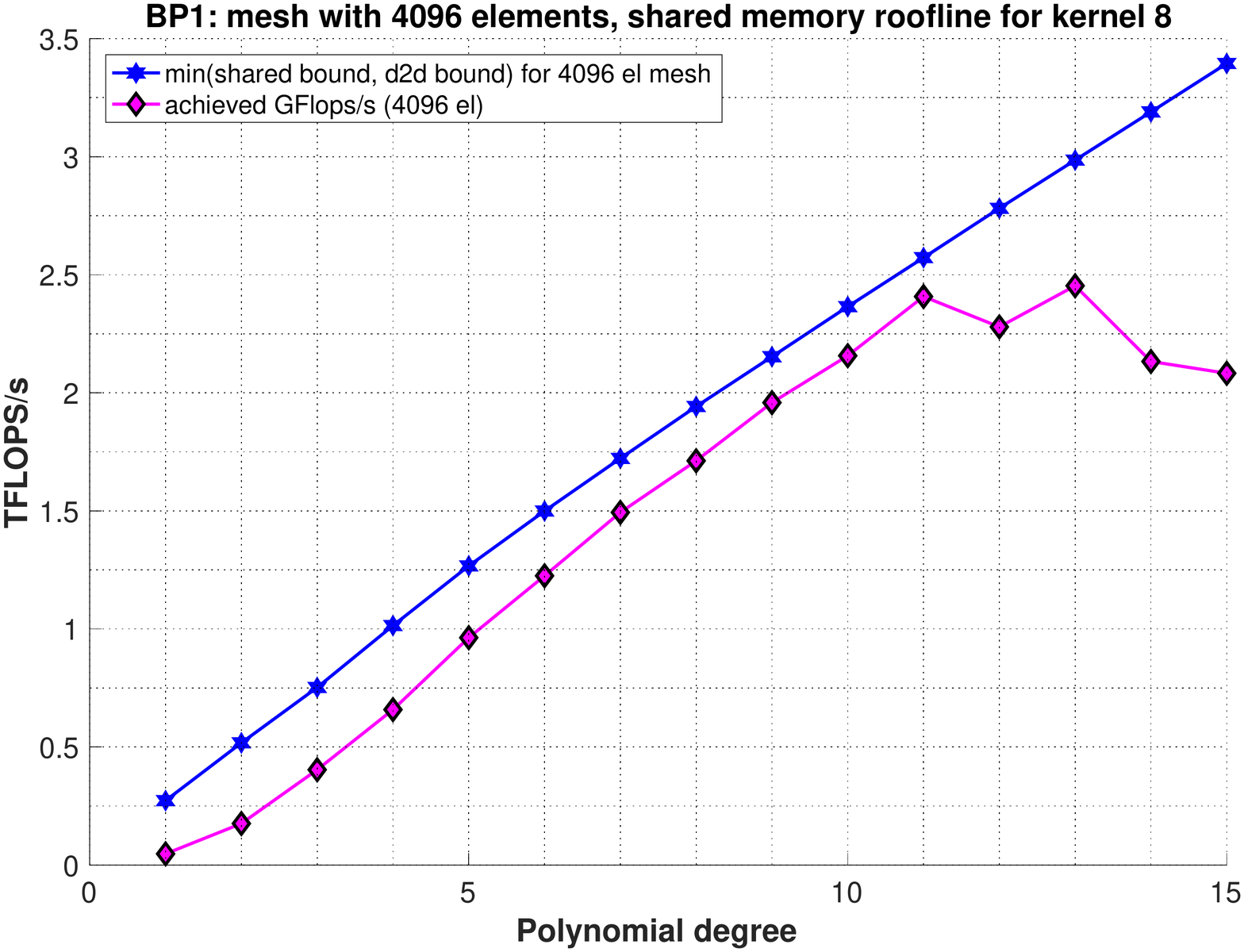}
\caption{BP1.0: shared memory rooflines and achieved floating point performance. Top left: kernel 6. Top right: kernel 7. Bottom: kernel 8.}
\label{fig:bp1sharedRoof}
\end{figure}

To explain the improvement between the Kernels 6-8 we consider the roofline model based on shared memory bandwidth. We generate a new roofline plot by taking a minimum of the empirical roofline based on device to device copy bandwidth \eqref{eq:globalRoof} and the upper limit computed based on shared memory bandwidth \eqref{eq:sharedRoof}. We show these new rooflines in Figure~\ref{fig:bp1sharedRoof}. For Kernel 8 this roofline is identical with the global memory roofline as in Figure~\ref{fig:bp1res} as the shared memory roofline is higher than the global memory roofline due to eliminating almost a half of shared memory transactions. This indicated that the shared memory bandwidth is not the limiting factor of the perforance of Kernel 8 for high $N$ where performance begins to degrade. For Kernels 6 and 7 however, we see that this shared memory roofline model indeed provides a reliable performance bound and these kernels are likely limited by shared memory performance. 

\section{Optimizing Benchmark Problem 3.5} \label{BP35.sec}

\noindent {\bf Kernel design.} As in BP1.0, we parallelize the action of the screened Poisson operator by assigning each element to a separate thread-block on the GPU. In this optimization procedure we investigate a 2D thread structure as well as test a 3D thread structure for $N<10$.

The action of the screened Poisson operator presents a difficult optimization challenge. In particular, hiding global memory latency is significantly more important than in BP1.0 because we transfer seven geometric factors ($\mathbf{J}^e$ and $6$ factors of $\mathbf{G}^e$) per node from global memory in every element. The majority of these values are required during the action of the stiffness matrix. To hide global memory latency, we aim to overlap data transfer with computations.

The increased amount of data that we are required to load increases both the number of global memory read transactions and the number of registers needed per thread. In theory, the geometric factors can be computed ``on the fly" inside the kernel using the element's vertices and the each node's $R_i,s_j,t_k)$ coordinates. This approach would reduced the amount of global memory loaded to $18+2N_q$ double precision values per block but increases the number of registers. We implemented and extensively tested this approach but concluded that it was impractical. Indeed, it is faster to simply load geometric factors from global memory than use extra registers and FLOPS. 

\noindent{\bf BP3.5 kernel optimization: 2D thread structure} We show the performance results of ten GPU kernels in Figure~\ref{fig:BP352D}. The ten kernels were constructed in sequence beginning from a direct implementation of the pseudo-code in Algorithm \ref{alg:bp35naive} and applying successive optimizations. The results is Figure~\ref{fig:BP352D} helps to demonstrate the effect each optimization has on the overall performance of BP3.5. As done with BP1.0, we list below some details of each kernel as well as any optimization made in that kernel. 

\begin{algorithm*}[t!]
\caption{BP3.5: starting point of the implementation (2D thread structure)}
  \label{alg:bp35naive}
\begin{boxedminipage}{\textwidth}
    \begin{algorithmic}[1]
     \STATE {{\bf Data:} (1) Vector $\mathbf{q}$, size $N_{el} \times N_p$, (2) differentiation matrix $\tilde{\mathbf{D}}^{\mathrm{1D}}$, size $N_q\times N_q$, (3) geometric factors $\mathbf{G}$, size $N_{el} \times N_p^{GL} \times 7$, (4)  parameter $\lambda$ };
    \STATE {{\bf Output: } Vector $\mathbf{Sq}$, size $N_{el} \times N_p$ };
    \FOR {$e\in \{1, 2, \ldots, N_{el}\}$}
    \FOR {each thread i,j}
    \STATE Load $\mathbf{D}^{\mathrm{1D}}$ to shared memory variable $\texttt{s}\_\texttt{D}$.  \COMMENT{Synchronize threads.}
    \STATE Allocate one shared auxiliary array \verb|s_tmps[Nq][Nq][Nq]| and three register arrays: \verb|r_Aq[Nq]|, \verb|r_qt[Nq]| and \verb|r_tmpt[Nq]| for storing intermediate results;
    \STATE $\verb|r_qt|_k = \sum_{n=1}^{N_q} \texttt{s}\_\texttt{D}_{kn}*\mathbf{q}_{kji}^{e}$;
 \FOR{$k\in \{1, 2, \ldots,N_q\}$}
    \STATE Load geometric factors to local variables \verb|G00|,  \verb|G01|,  \verb|G02|,  \verb|G11|,  \verb|G12|,  \verb|G12|,  \verb|G22|, \verb|GwJ|
    \STATE Declare variables \verb|qr|  and \verb|qs| and set them to \verb|0|.
    	\STATE \verb|qr| = $\sum_{n=1}^{N_q} \texttt{s}\_\texttt{D}_{in}*\mathbf{q}_{kjn}^{e}$;
		\STATE \verb|qs |= $\sum_{n=1}^{N_q} \texttt{s}\_\texttt{D}_{jn}*\mathbf{q}_{kni}^{e}$;
    \STATE \verb|Sqtemp|$_{kji}$ = \verb|G00*qr| + \verb|G01*qs| + \verb|G02*r_qt|$_k$;
	\STATE \verb|s_tmps|$_{kji}$ = \verb|G01*qr| + \verb|G11*qs| + \verb|G12*r_qt|$_k$;
	\STATE \verb|r_tmpt|$_k$ = \verb|G02*qr| + \verb|G12*qs| + \verb|G22*r_qt|$_k$;

	\STATE $\mathbf{Sq}_{kji}^{e} = \mathbf{Sq}_{kji}^{e}$+\verb|GwJ|*$\lambda$*$\mathbf{q}_{kji}^{e}$;
	\STATE $\mathbf{Sq}_{kji}^{e} = \sum_{n=1}^{N_q}; \mathbf{Sq}_{nji}^{e}$*\verb|r_tmpt|$_k$;
	 \ENDFOR\COMMENT{Synchronize threads.}
 \FOR{$k \in \{1,2, \ldots, N_q\}$}
  \STATE Declare variables \verb|Sq1|  and \verb|Sq2| and set them to \verb|0|.
  \STATE	\verb|Sq1| = $\sum_{n=1}^{N_q}$\verb|Sqtemp|$_{kjn}*\texttt{s}\_\texttt{D}_{ni}$;
\STATE		\verb|Sq2| = $\sum_{n=1}^{N_q}$\verb|Sqtemp|$_{kni}*\texttt{s}\_\texttt{D}_{nj}$;

  \STATE $\mathbf{Sq}_{kni}^{e} = \mathbf{Sq}_{kni}^{e}+ 	\verb|Sq1| + 	\verb|Sq2|;$ 
 \ENDFOR\COMMENT{Synchronize threads.}
    \ENDFOR 
    \ENDFOR
  \end{algorithmic}
\end{boxedminipage}
\end{algorithm*}

\begin{figure}[t]
\centering
\includegraphics[width=0.85\textwidth]{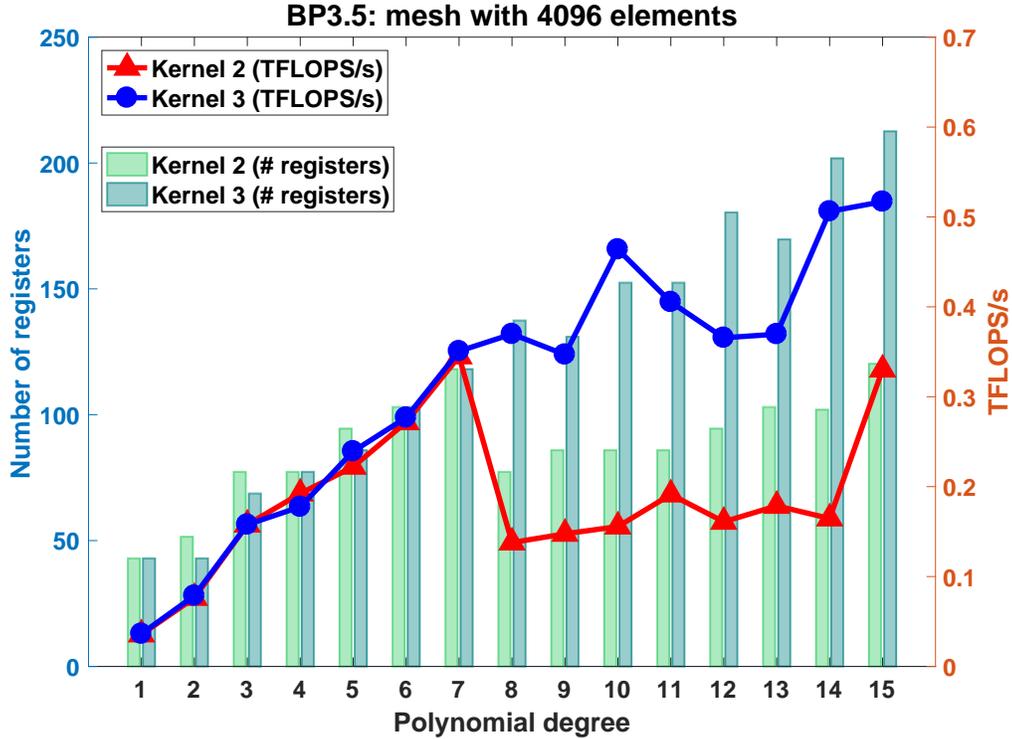} 
\caption{BP3.5: Comparison of performance for Kernel 1 and Kernel 2. The only difference between these two kernels is loop unrolling. The number of registers per thread is shown in the bar chart in the background.}
\label{fig:bp35unroll}
\end{figure}

\begin{figure*}[t]
\centering
\includegraphics[width=0.45\textwidth]{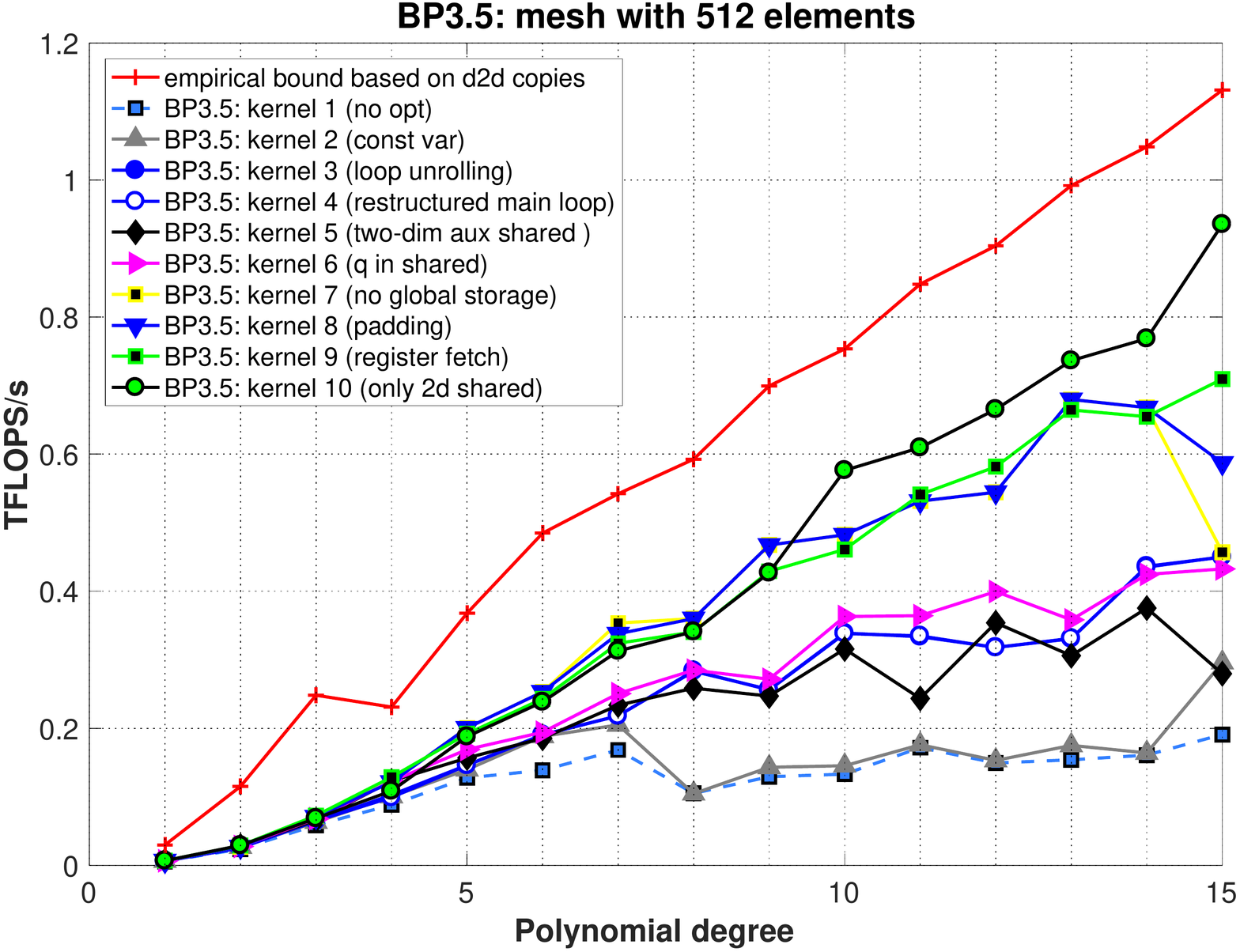} \includegraphics[width=0.45\textwidth]{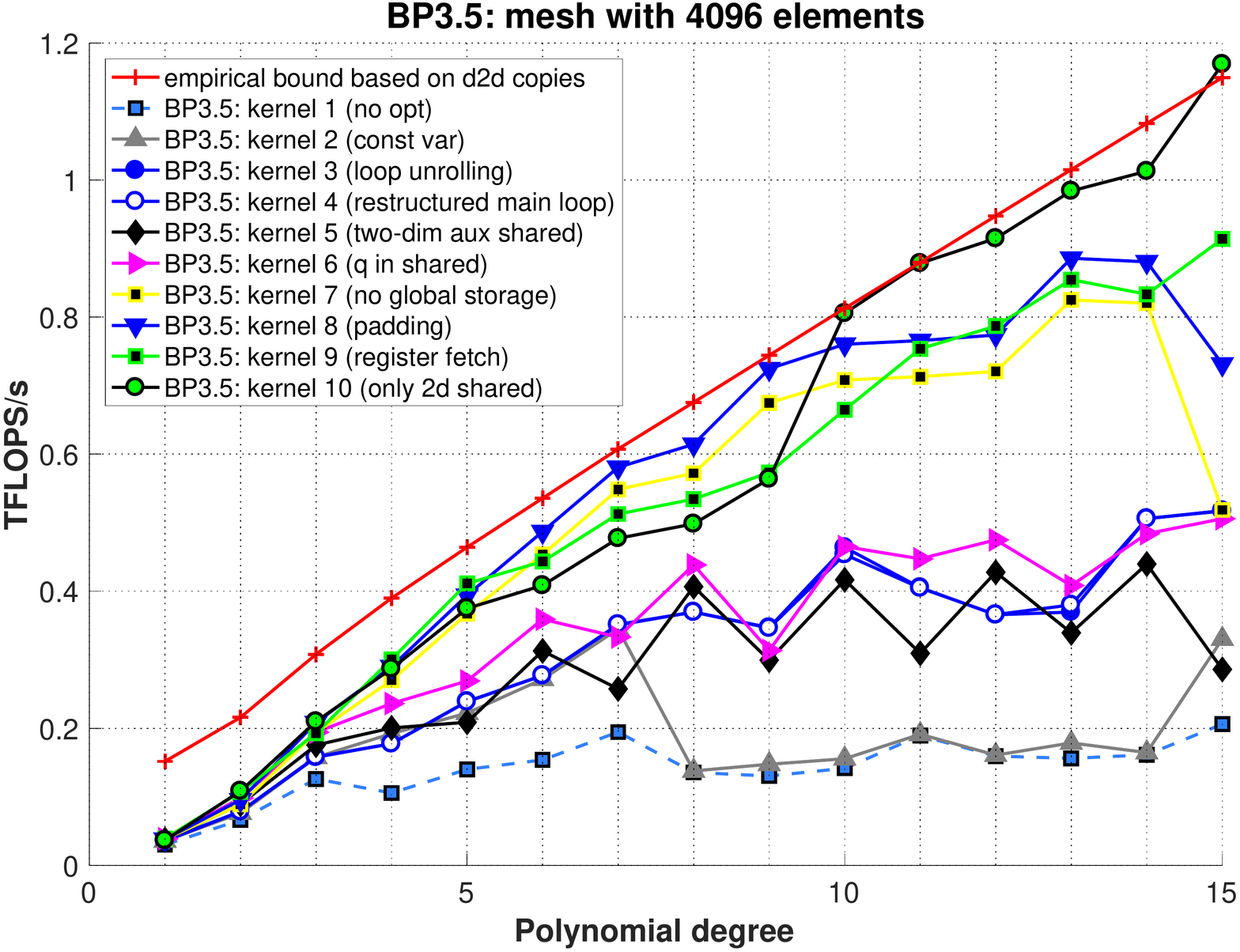}
\caption{BP3.5: Performance of 2D thread array kernels in various stages of optimization. The red line marked with crosses is the empirically determined roofline based on  achievable device to device memory copies on an NVIDIA P100 PCI-E 12GB GPU. Left: TFLOPS/s for cubical mesh with $512$ elements. Right: TFLOPS/s for cubical mesh with $4096$ elements.}
\label{fig:BP352D}
\end{figure*}

\kernelfl{1} This kernel is a direct implementation of Algorithm~\ref{alg:bp35naive}. The one-dimensional differentiation matrix $\mathbf{D}^{\mathrm{1D}}$ is fetched to shared memory at the beginning of the kernel. This kernel also uses a shared memory array of size $N_p$ to store partial sums. Entries of $\mathbf{q}^{e}$ are fetched from global memory as needed. In the last loop, the partial sums are successively added directly to the global memory array to store the final result. The kernel achieves 200 GFLOPS/s, which is one sixth of the predicted empirical roofline.
\kernel{2} In this kernel all the variables, except the array used for storing the final result, are declared using \verb|const|. This optimization has only a marginal influence on the performance for $N\geq 8$.
\kernel{3} In this kernel all loops with $k$ are unrolled. Unrolling loops improves the performance for the larger mesh only if $N\geq 8$ and for all $N$ for the smaller mesh. Note that a kernel with unrolled loops uses more registers per thread and, hence, the achieved occupancy decreases. The measured TFLOPS/s increases, however, as shown in Figure~\ref{fig:bp35unroll}. This behavior can be explained if unrolling has increased instruction-level parallelism.
\kernel{4} In this kernel we place the $k$ loop (lines 7--17 in Algorithm \ref{alg:bp35naive}) on the exterior of $i$ and $j$ loops. Doing this, $k$ becomes the slowest running index. This loop structure is justified because we iterate through $(i,j)$ slices.
\kernel{5} In this kernel the auxiliary shared memory array of size $N_p$ is replaced by two auxiliary shared memory array of size $N_q^2$ each. 
\kernel{6} In this kernel the field variable $\mathbf{q}^{e}$ is fetched to the shared memory. The overall improvement is modest and the performance reaches 0.5 TFLOP/s at this point. 
\kernel{7} This kernel reduces total global memory transactions by caching the necessary data at the beginning of the kernel and only writing the output variable once. This produces a significant optimization and improves performance by about $40\%$.
\kernel{8} For this kernel, if $N=7$ or $15$, the arrays are padded to avoid shared memory bank conflicts. The improvements are significant for the larger mesh and $N=15$.
\kernel{9} In this kernel each thread allocates a register array with $N_q$ elements and the field variable $\mathbf{q}^{e}$ is fetched to these registers instead of shared memory. For both meshes, this approach slightly improves the performance, and only for $N=12$ and $N=15$.
\kernel{10} This kernel uses three two-dimensional shared memory arrays for partial results and fetches the field variable $\mathbf{q}^{e}$ to register arrays. The achieved TFLOPS/s for this kernel are aligned with the empirical roofline.

\noindent{\bf BP3.5 results: 2D thread structure.} Many of the performance improvements we observe when optimizing the kernels for BP3.5 are due to subtle changes in the code, e.g. declaring variables as constant, adding padding, or unrolling the loops. But global and shared memory usage has the biggest impact on the performance. Once we reduce the use of global memory (starting in Kernel 7) the performance improves substantially, especially for larger $N$. Reducing the amount of shared memory and caching the data to registers results in the second most important factor which is visible in Figure~\ref{fig:BP352D} for the mesh with $4096$ elements and $N\geq 10$.

For the reference kernel, Kernel 1, the performance barely reaches $200$ GFLOPS/s whereas the most optimized kernel we present, Kernel 10, achieved up to $1.2$ TFLOPS/s. While this is only a six fold speedup, comparison with our empirical roofline model, based on streaming global device memory, shows that the performance of Kernel 10 is comparable to just streaming the minimally necessary data.


\begin{algorithm*}[t!]
  \caption{BP3.5: collocation differentiation for 3D hexahedral mesh (3D thread structure)}
  \label{alg:bp35pseudocode3D}
\begin{boxedminipage}{\textwidth}
{\small
    \begin{algorithmic}[1]
      \STATE {{\bf Data:} (1) Vector $\mathbf{q}$, size $N_{el} \times N_p$, (2) differentiation matrix $\tilde{\mathbf{D}}^{\mathrm{1D}}$, size $N_q\times N_q$, (3) geometric factors $\mathbf{G}$, size $N_{el} \times N_p^{GL} \times 7$, (4)  parameter $\lambda$ };
    \STATE {{\bf Output: } Vector $\mathbf{Sq}$, size $N_{el} \times N_p$ };
      \FOR {For $e\in \{1,2, \ldots, N_{el}\}$}
    \FOR {$i, j, k \in \{1, 2, \ldots N_q\}$}
    \STATE If \verb|k=0|, load $\mathbf{D}^{\mathrm{1D}}$ to shared memory variable $\texttt{s}\_\texttt{D}$;
    \STATE Declare register variables \verb|r_qr|, \verb|r_qs|, and \verb|r_qt|;
    \ENDFOR \COMMENT{Synchronize threads.}
      \FOR {$i, j, k \in \{1, 2, \ldots N_q\}$}
    \STATE Load \verb|GwJ|;
     \STATE Declare variables \verb|qr|, \verb|qs|, \verb|qt| and set them to \verb|0|.
    	  \STATE  \verb|qr| = $\sum_{n=1}^{N_q} \texttt{s}\_\texttt{D}_{in}*\mathbf{q}_{kjn}^e$;
		\STATE	  \verb|qs| = $\sum_{n=1}^{N_q} \texttt{s}\_\texttt{D}_{jn}*\mathbf{q}_{kni}^e$;
		\STATE	  \verb|qt| = $\sum_{n=1}^{N_q} \texttt{s}\_\texttt{D}_{kn}*\mathbf{q}_{nji}^e$;
    
    \STATE Set \verb|r_qr = qr|, \verb|r_qs = qs| and \verb|r_qt = qt|;
    \STATE  $\mathbf{Sq}_{kji}^e$ =\verb|GwJ|*$\lambda*\mathbf{q}_{kji}^e$;
   \ENDFOR\COMMENT{Synchronize threads.}
      \FOR {$i, j, k \in \{1, 2, \ldots N_q\}$}
     \STATE Load \verb|G00|, \verb|G01|, \verb|G02|;
     \STATE \verb|Sqtemp|$_{kji}$= \verb|G00*r_qr + G01*r_qs + G02*r_qt|;
     \ENDFOR
      \FOR {$i, j, k \in \{1, 2, \ldots N_q\}$}
		   \STATE  $\mathbf{Sq}_{kji}^e$ = $\mathbf{Sq}_{kji}^e+\sum_{n=1}^{N_q}\texttt{s}\_\texttt{D}_{ni}* \texttt{Sqtemp}_{kjn}$;
     \ENDFOR\COMMENT{Synchronize threads.}
         \FOR {$i, j, k \in \{1, 2, \ldots N_q\}$}
     \STATE Load \verb|G10|, \verb|G11|, \verb|G12|;
     \STATE \verb|Sqtemp|$_{kji}$ = \verb|G10*r_qr + G11*r_qs + G12*r_qt|;
     \ENDFOR\COMMENT{Synchronize threads.}
      \FOR {$i, j, k \in \{1, 2, \ldots N_q\}$}
		   \STATE  $\mathbf{Sq}_{kji}^e$ = $\mathbf{Sq}_{kji}^e+\sum_{n=1}^{N_q}\texttt{s}\_\texttt{D}_{nj}* \texttt{Sqtemp}_{kni}$;
     \ENDFOR\COMMENT{Synchronize threads.}
     
         \FOR {$i, j, k \in \{1, 2, \ldots N_q\}$}
     \STATE Load \verb|G10|, \verb|G11|, \verb|G12|;
     \STATE \verb|Sqtemp|$_{kji}$ = \verb| G20*r_qr + G21*r_qs + G22*r_qt|;
     \ENDFOR\COMMENT{Synchronize threads.}
      \FOR {$i, j, k \in \{1, 2, \ldots N_q\}$}
		   \STATE $\mathbf{Sq}_{kji}^e$ = $\mathbf{Sq}_{kji}^e+\sum_{n=1}^{N_q}\texttt{s}\_\texttt{D}_{nk}* \texttt{Sqtemp}_{nji}$;
     \ENDFOR
    \ENDFOR
    
\end{algorithmic}
}
\end{boxedminipage}
\end{algorithm*}

\noindent{\bf BP3.5 kernel optimization: 3D thread structure}. We show the performance results of six GPU kernels in Figure~\ref{fig:BP352D}. We again construct these kernels as a sequence of successive optimizations beginning from the the pseudo-code shown in Algorithm~\ref{alg:bp35pseudocode3D} which uses a 3D thread structure. we present results for these kernels for $N=1, 2,   \ldots, 9$, since associating one thread with one node as done in this 3D thread strructure would require more than the maximum of 1024 threads for higher-degree polynomials. We list below some details of each kernel as well as any optimization made in that kernel.

\kernelfl{1} This kernel serves as an initial reference kernel and is a direct implementation of Algorithm \ref{alg:bp35pseudocode3D}. The differentiation matrix $\mathbf{D}^{\mathrm{1D}}$ is loaded into shared memory and partial sums are stored in global arrays. The performance of this kernel reaches 300 GFLOPS/s in the best case. 
\kernel{2} In this kernel all variables, except the output variable, are declared using \verb|const|. All loops with index $n$ are unrolled. The overall improvement is modest. 
\kernel{3} In this kernel the geometric factors are stored in registers and fetched only once from global memory. This reduces the number of global memory loads from nine to seven. This kernel reaches about 350 GFLOPS/s for the larger mesh. In case of the smaller mesh, performance of this kernel is similar to Kernel 2. 
\kernel{4} Instead of writing the result directly to \verb|Aq|, in this kernel the partial results are accumulated in a register variable. The performance of this kernel is only slightly better than that of Kernel 3.
\kernel{5} In this kernel the field variable $\mathbf{q}^{e}$ is cached to shared memory. The same shared array is used to store the partial result, hence there is no need to use \verb|Sqtemp|. The performance increases significantly and reaches 610 GFLOPS/s.
\kernel{6} In this kernel, partial results are stored in three shared memory arrays of size $N_p$ each. This strategy reduces occupancy but allows the kernel to use less synchronizations (the number of synchronizations reduced to two). However, there is almost no performance difference between this and the previous kernel due to extensive use of shared memory. 

\noindent{\bf BP3.5 results: 3D thread structure.} 

The kernels adapting 3D thread structure form two groups. First group consists of Kernels 1--4 and the second one consists of Kernels 5 and 6. The largest performance jump appears between Kernels 4 and 5. It is a result of caching the field variable to shared memory. Kernel 5 appears to be the best performing kernel. As a result of succesively reducing the number of shared memory fetches, we improved performance by a factor of two, from approximately 300 GFLOPS/s to around 600 GFLOPS/s.

\begin{figure*}[t]
\centering
\includegraphics[width=0.45\textwidth]{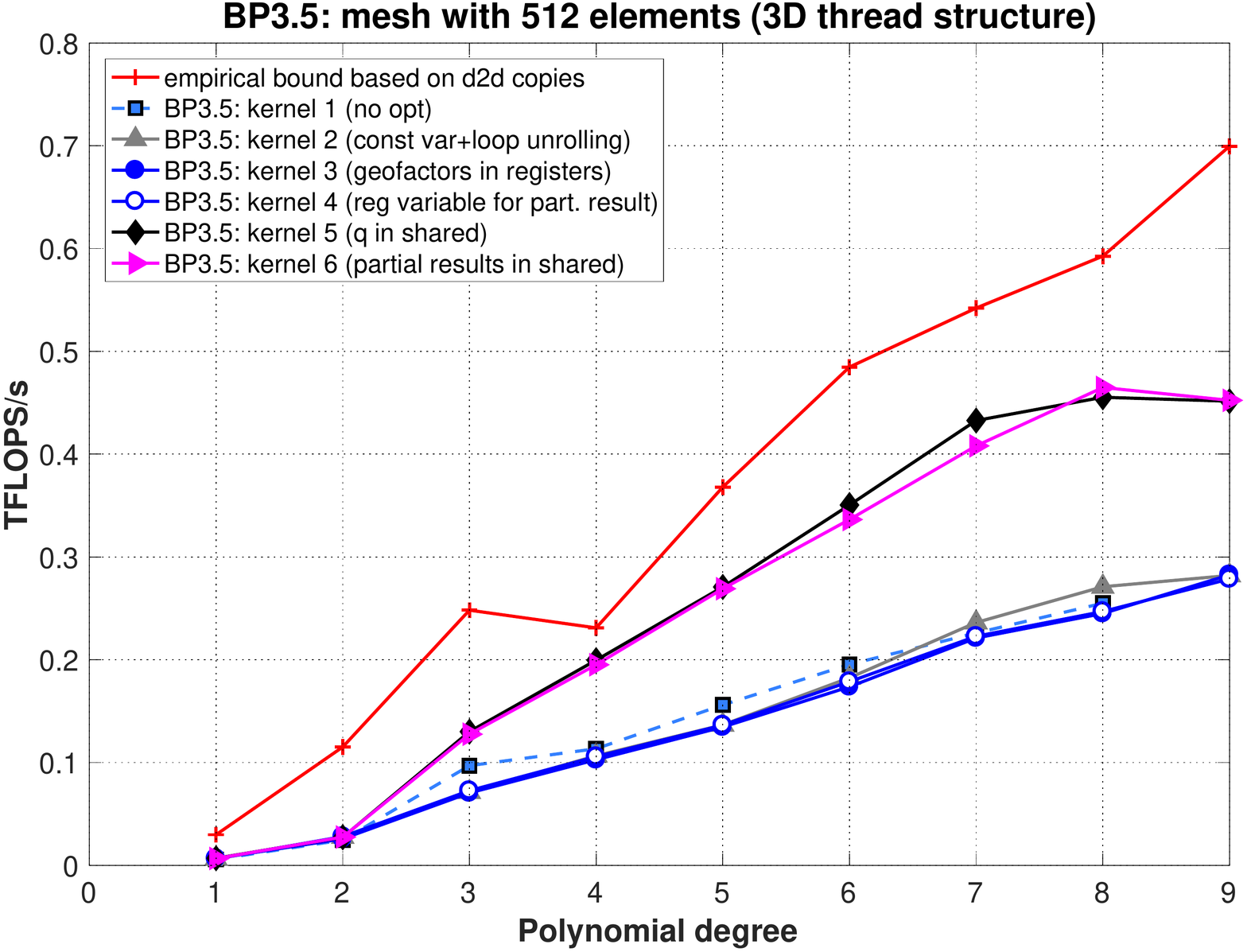} \includegraphics[width=0.45\textwidth]{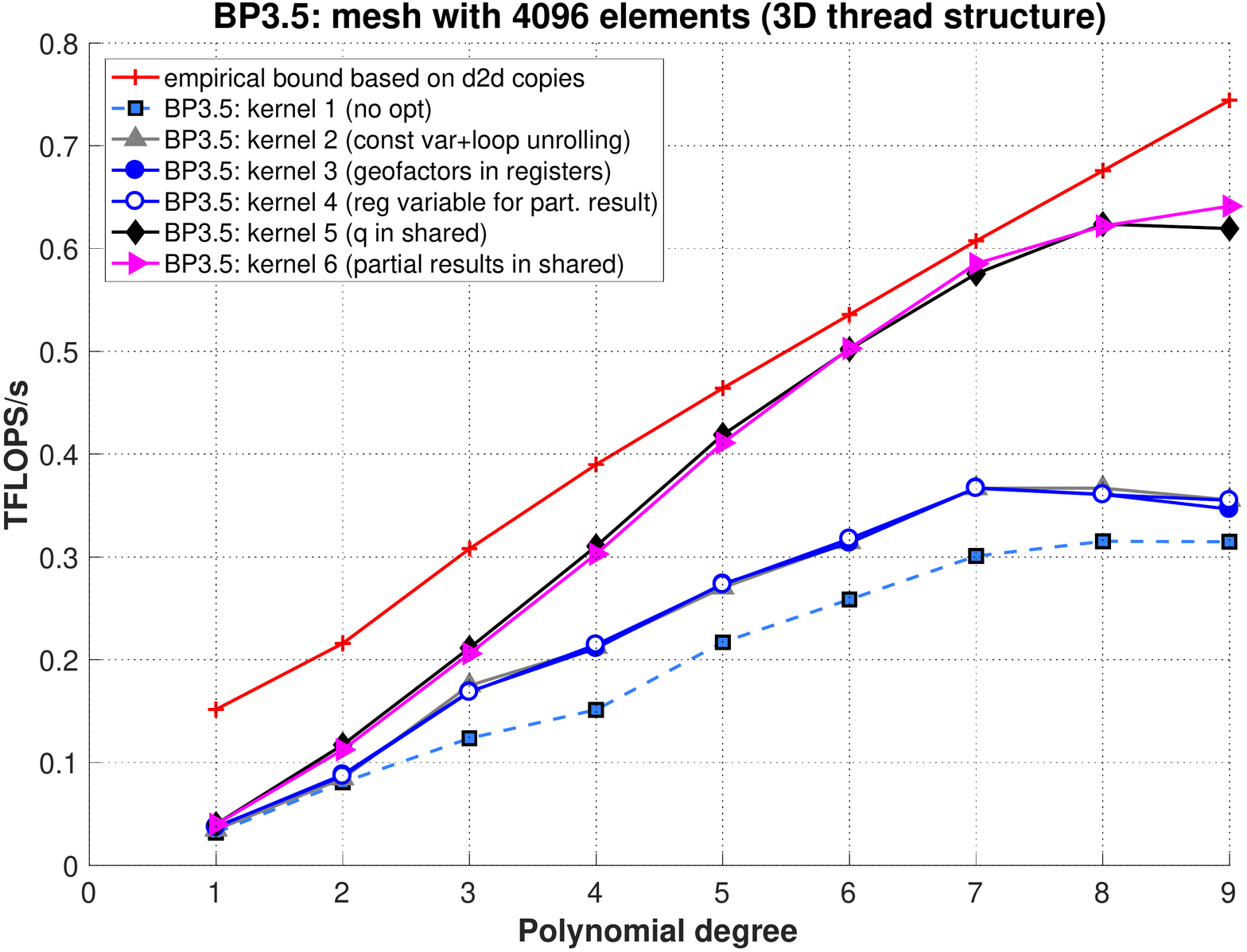}
\caption{BP3.5: Performance of 3D kernels in various stages of optimization. The red line marked with crosses is the roofline computed based on device to device copies measured on an NVIDIA P100 PCI-E 12GB GPU. Left: TFLOPS/s for cubical mesh with $512$ elements. Right: TFLOPS/s for cubical mesh with $4096$ elements.}
\label{fig:BP353D}
\end{figure*}

\section{Optimizing Benchmark Problem 3.0} \label{BP3.sec}

\noindent {\bf BP3.0 Kernel Design.} BP3.0 can be considered a fusion of BP1.0 and BP3.5. This benchmark shares its interpolation/projection steps with BP1.0 and the stiffness matrix action with BP3.5. Hence, BP3.0 inherits all the optimization challenges we have discussed thus far for BP1.0 and BP3.5. In particular, the kernel for this problem needs to synchronize threads multiple times and needs to load a large amount of data from global memory. Furthermore, compared to either BP1.0 or BP3.5 this benchmark requires even more shared memory usage in order to store partial results. Since differentiation is performed using a denser set of nodes, we also transfer more data per thread block compared with BP3.5. 

The action of the screened Poisson operator in this benchmark can be written as three distinct parts: interpolation to GL nodes, stiffness and mass matrix actions, and projection back to GLL nodes. Therefore it is possible to split the implementation into three kernels. This approach reduces memory requirements per kernel and makes the code more readable. We investigate this approach with 3D thread structure.

\begin{figure*}[t]
\centering
\includegraphics[width=0.45\textwidth]{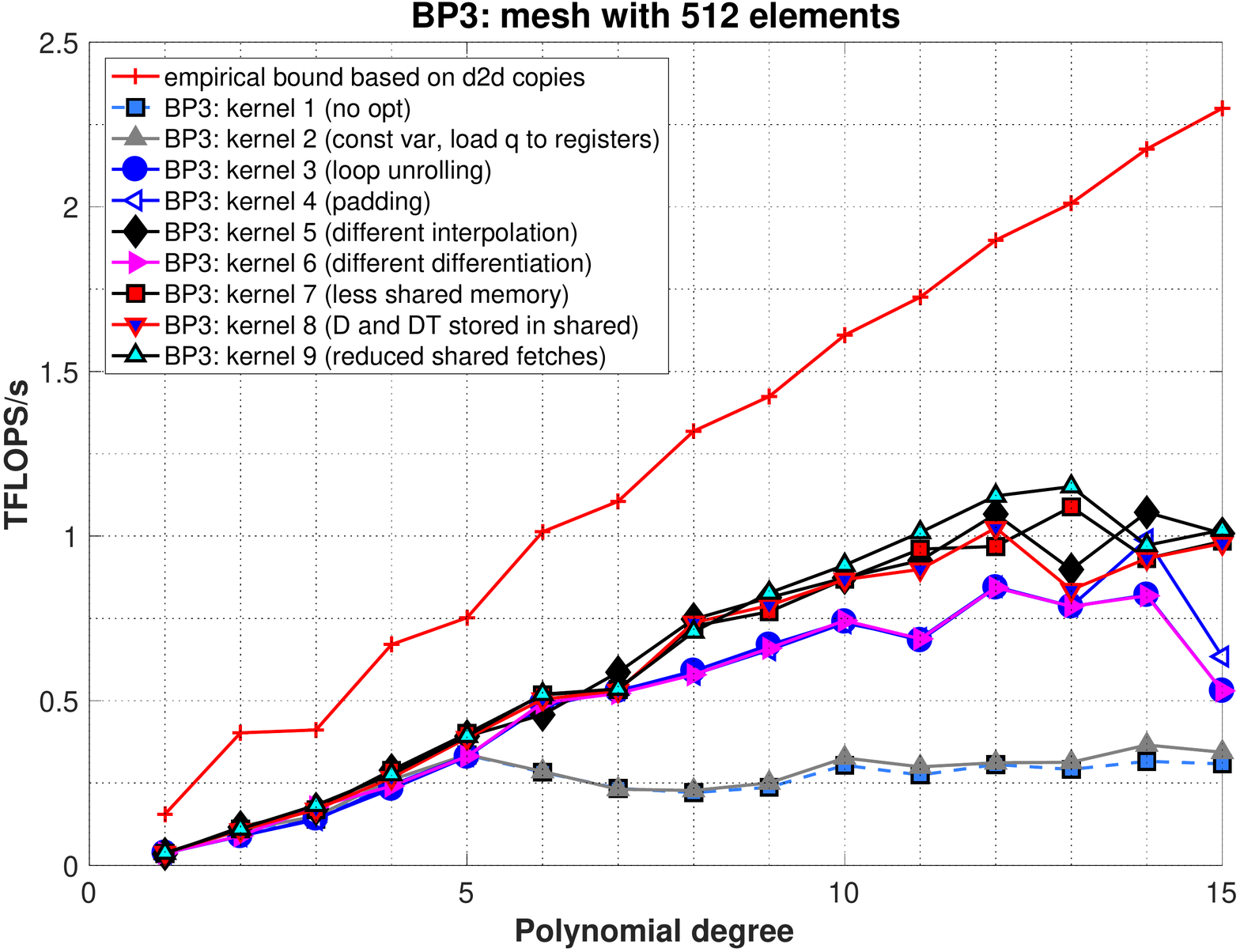} \includegraphics[width=0.45\textwidth]{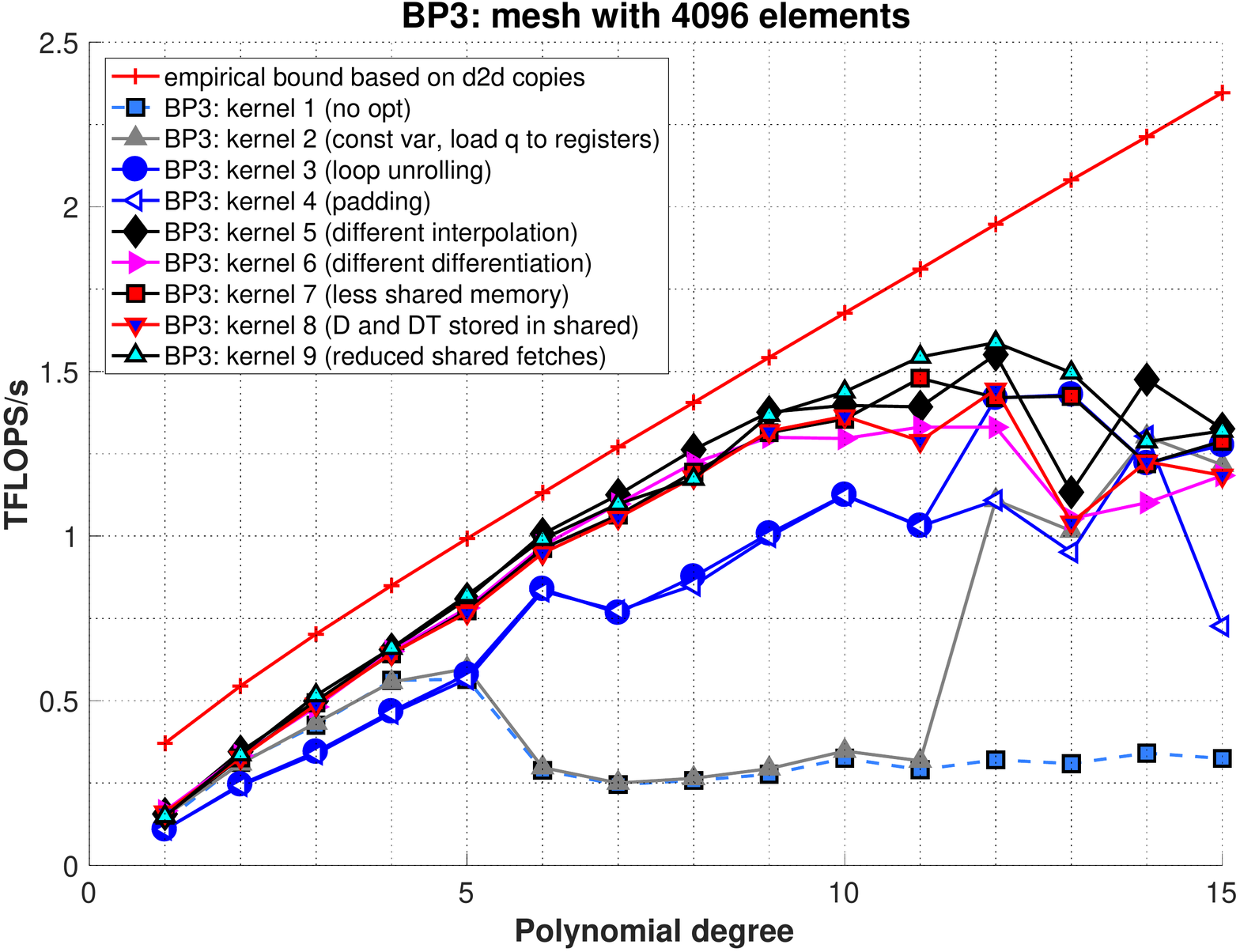}
\caption{BP3.0: Performance of 2D thread array kernels in various stages of optimization. The red line marked with crosses is the roofline computed based on device to device copies on a single NVIDIA P100 PCI-E 12GB GPU. Left: TFLOPS/s for cubical mesh with $512$ elements. Right: TFLOPS/s for cubical mesh with $4096$ elements. }
\label{fig:BP3GFLOPS}
\end{figure*}

\noindent {\bf BP3.0 kernel optimization: 2D thread structure.} We begin as in BP3.5 by first investigating a kernel implementation using a 2D thread structure. The performance results for nine separate kernel are presented in Figure~\ref{fig:BP3GFLOPS}. As in the previous benchmarks we present a sequence of kernels and detail what optimizations we performed in each kernel. 

\kernelfl{1} This kernel serves as a reference implementation. In this kernel there are four shared memory arrays: one for  $\mathbf{I}^{\mathrm{1D}}$, one for $\mathbf{D}^{\mathrm{1D}}$ and two arrays with $\left(N_q^{GL}\right)^2$ elements for storing partial sums. Each thread also allocates an additional register array to store additional partial sums that do not require sharing between threads in a thread block. The field variable $\mathbf{q}^{e}$ is read directly from global memory when needed. This kernel reaches approximately 280 GFFLOPS/s.
\kernel{2} In this kernel all the input variables, except the pointer used for storing the final results, are declared as \verb|const|. Each thread allocates a register array of size $N_q$ and caches a column of entries of $\mathbf{q}^{e}$ to the register array. The performance for $N\geq 12$ for the larger mesh improves to over 1.25 TFLOP/s
\kernel{3} In this kernel all internal loops are unrolled. The performance improves to 600 GFLOPS/s for the smaller mesh and to 1.4 TFLOP/s for the larger mesh.
\kernel{4} In this kernel we add padding to all the shared memory arrays to avoid bank conflicts. In particular, in case $N_q^{GL}=8$ or $N_q^{GL}=16$, the arrays with size $N_q^{GL}\times N_q^{GL}$ change their size to $N_q^{GL}\times \left(N_q^{GL}+1\right)$.  Padding helps only for the smaller mesh.
\kernel{5} This kernel adapts the most efficient interpolation strategy devised for BP1.0 (see BP1.0: Kernel 7 and Figure 4). For the larger mesh, the performance reaches 1.5 TFLOP/s.
\kernel{6} In this kernel the field variable $\mathbf{q}^{e}$ is loaded to a shared memory array. This kernel implements more efficient differentiation (as in BP3.5: Kernel 9). The improvement is very modest since in order to use this type of differentiation we need to combine it with less efficient interpolation. 
\kernel{7} A version of Kernel 6 with one less shared memory array. We use only one shared memory array to store the partial result at a cost of additional synchronizations. Kernel 7 performs better than Kernel 6, however in most cases, it is not more efficient than Kernel 5.
\kernel{8} This kernel is a version of Kernel 7 with both $\mathbf{D}^{\mathrm{1D}}$ and $(\mathbf{D}^{\mathrm{1D}})^T$ fetched to shared memory. Now all the threads can access shared memory column-wise. Performance is very similar to Kernel 6.
\kernel{9} The kernel uses the same strategy as in the last kernel implemented for BP1.0, i.e., it fetches each repeating entry of the interpolation matrix $\mathbf{I}^{\mathrm{1D}}$ only once. The performance of this kernel is the best that we present here as it reaches 1.6 TFLOPS/s.



\noindent {\bf BP3.0 Results: 2D thread structure.} For this benchmark problem, our best performing kernel for high-order FEM approximations, Kernel 10, performs approximately four times as the reference kernel, Kernel 1. Although we did not achieve the peak performance as predicted by our empirical roofline model using global memory bandwidth, for the 4096 element mesh and $N\leq 12$ our kernels are very close to that peak. 

\begin{figure}[t]
\centering
\includegraphics[width=0.45\textwidth]{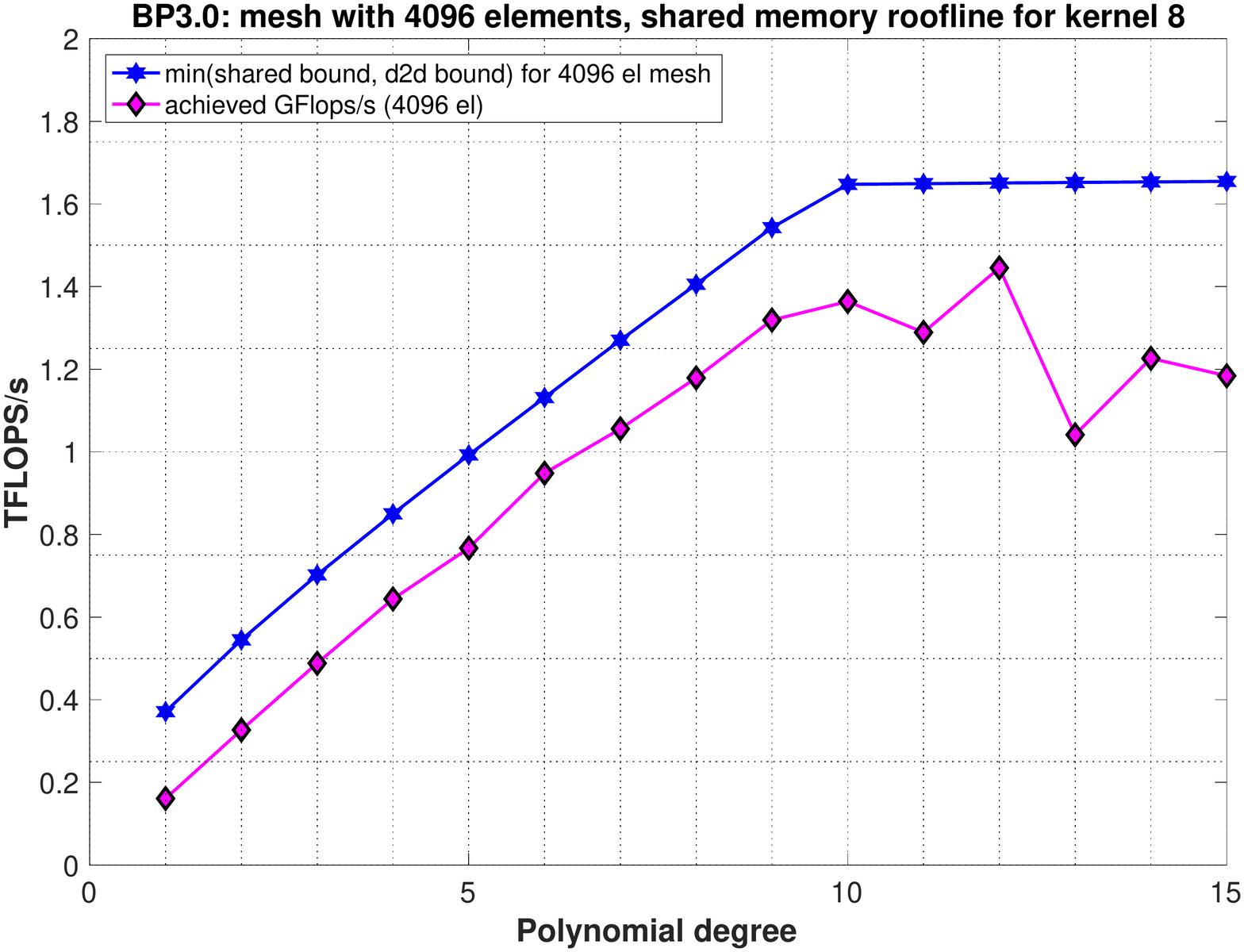}\includegraphics[width=0.45\textwidth]{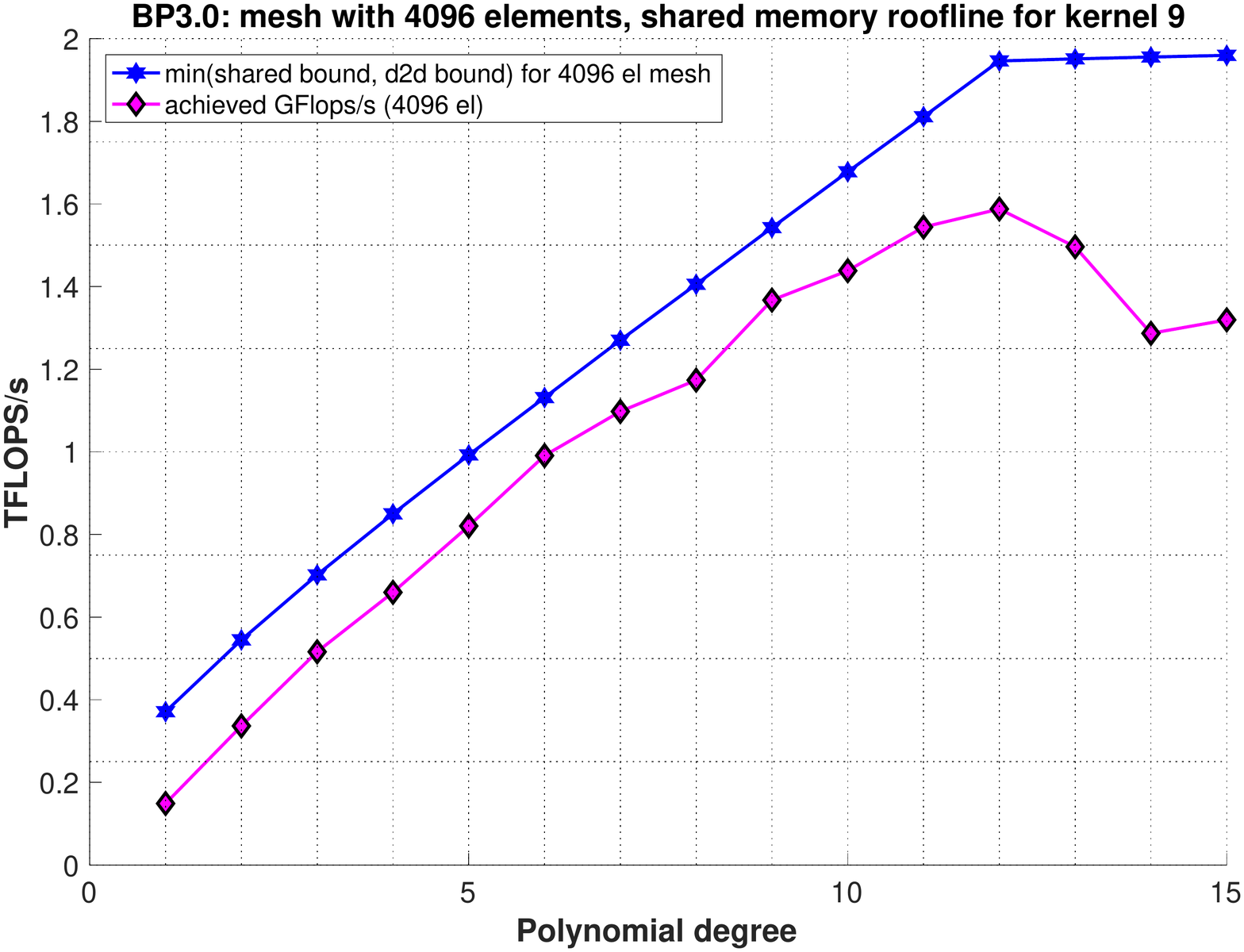}
\caption{BP3.0: lines for kernels 8 and 9. Top: achieved floating point performance and shared memory roofline for Kernel 8. Bottom: achieved floating point performance and shared memory roofline for Kernel 9.}
\label{fig:bp3sharedRoof}
\end{figure}

As in BP1.0 we can obtain a better understanding of the performance of our kernels by considering the roofline model based on shared memory bandwidth. Figure~\ref{fig:bp3sharedRoof} shows updated rooflines for Kernels 8 and 9 in which we see that this shared memory bandwidth bound yields a better estimate for the achieved performance of these kernels with $N\geq10$.

\noindent {\bf BP3.0 3D thread structure.}  We perform analogous tuning process using a 3D thread structure for $N<9$. Figure~\ref{fig:BP33DGFLOPS} shows the performance of these kernels presented below.
\kernel{1} This kernel starts from a direct implementation of the psuedo-code given in Algorithm \ref{alg:bp3pseudocode}, which is executed as three separate CUDA kernels. The field variable $\mathbf{q}^{e}$ is read directly from global memory and partial results are stored in global memory as well. The code reaches approximately 280 GFLOPS/s in the best case.
\kernel{2} This kernel combines the three separate kernels as one CUDA kernel. There is almost no difference between Kernel 1 and Kernel 2. This suggests that the cost of additional kernel launches is small compared with the cost of data transfer.
\kernel{3} In this kernel all  internal loops are unrolled. Unrolling loops would appear to be more important for 2D thread structure as for 3D thread structure it has only a marginal influence on performance.
\kernel{4} This kernel differs from Kernel 2 by loading the geometric factors once and avoiding redundant reads. The gain in performance is small because of large amount of global memory transactions that have not yet been eliminated.
\kernel{5} While we access the field variable $\mathbf{q}^{e}$ directly from global memory in the interpolation and projection parts, $\mathbf{q}^{e}$ is fetched to a shared array in the differentiation step. 
\kernel{6}  This kernel is a version of Kernel 5 with intermediate results stored in registers in the differentiation step. Kernels 5 and 6 are approximately 1.5 times faster than Kernel 4.
\kernel{7} In this kernel we store the partial results in shared memory and registers everywhere except between interpolation, first and second differentiation, and projection.
\kernel{8} In this kernel there is no intermediate global storage. All intermediate results are accumulated either in shared memory or in registers. 
\kernel{9} In this kernel we declare three additional shared memory arrays (size $N_p$ each) for storing intermediate results. As a result, the performance reaches 900 GFLOPS/s.

\noindent {\bf BP3.0 Results: 3D thread structure.} Unlike in the case of BP3.5, in BP3.0 we observe a very gradual improvement in performance. However, as for BP3.0 the largest improvements appear when removing additional global memory transactions. It is interesting to note that the difference between kernel 7 and 8 is rather small, despite the fact that Kernel 7 reads and writes three vectors of size $N_{el}\cdot N_p$ to the global memory and Kernel 8 reads and writes only once. The lack of significant improvement may be a result of the additonal synchronizations enforced in Kernel 8. In Kernel 7, the global write serves as synchronization. 


\begin{figure*}[t]
\centering
\includegraphics[width=0.45\textwidth]{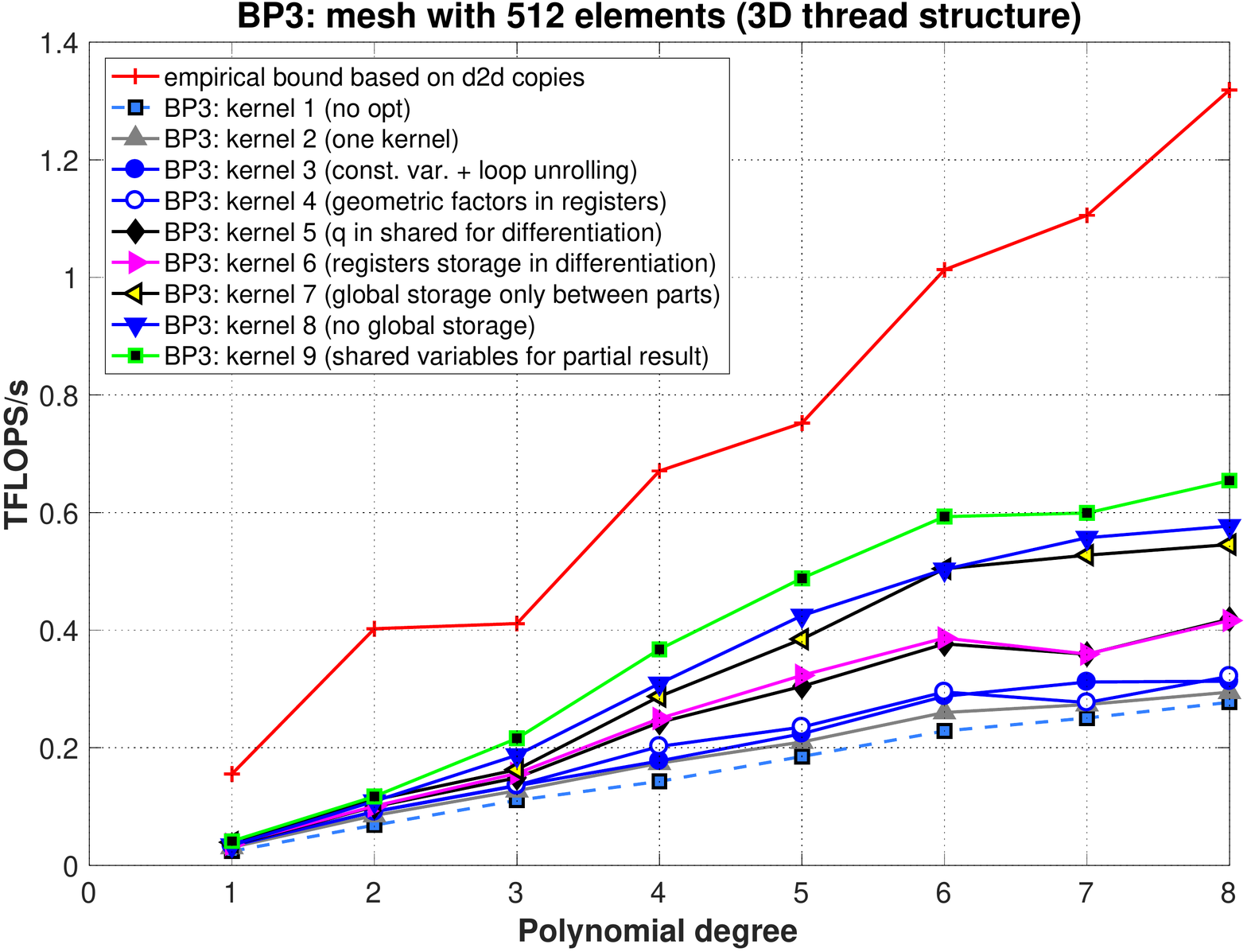} \includegraphics[width=0.45\textwidth]{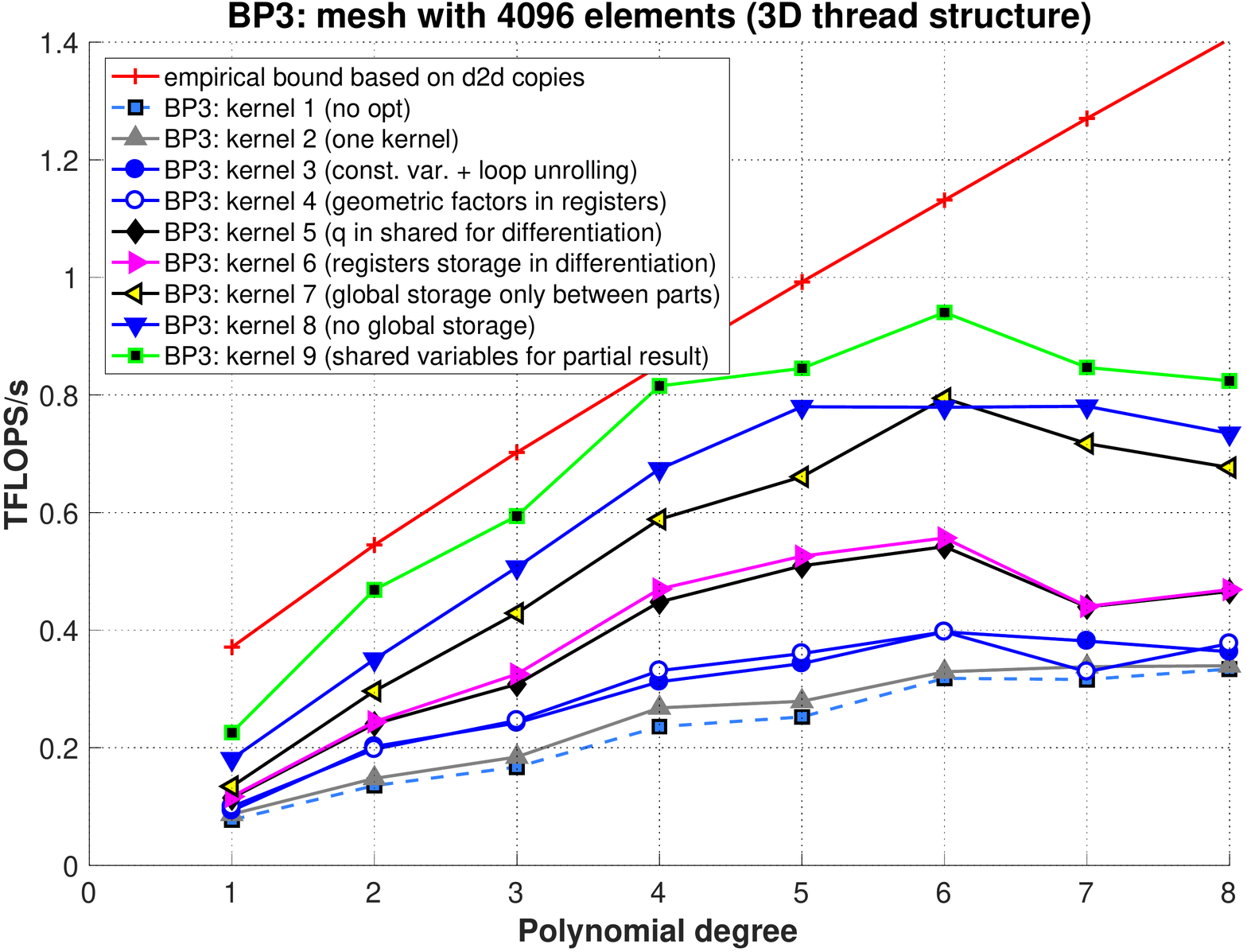}
\caption{BP3.0: Performance of 3D kernels in various stages of optimization. The red line marked with crosses is the roofline computed based on device to device copies on a single NVIDIA P100 PCI-E 12GB GPU. Left: TFLOPS/s for cubical mesh with $512$ elements. Right: TFLOPS/s for cubical mesh with $4096$ elements. Note: almost all kernels fail for $N=3$ on the smaller mesh.}
\label{fig:BP33DGFLOPS}
\end{figure*}

\section{Conclusions and future work}
\label{conclusions.sec}

In this paper we described the GPU optimization of three FEM-specific benchmark problems which have significant relevance in real world large scale code bases. We detailed how reducing factors such as global memory transactions and reducing pressure on shared memory usage can lead to efficient GPU kernels for these FEM operators. For BP3.5, the most tuned kernel is aligned with the empirical roofline bound and its performance is limited only by global memory bandwidth. We obtained similar results for the two remaining problems but their performance does not scale perfectly for $N\geq 13$. For each of the problems we used a sequence of optimizations. One might argue that multiple authors (see, for example \citep{Garvey2015}, \citep{Abdelfattah2016}, \citep{Nelson2015}) consider automatic performance tuning and our efforts should rather be replaced by a computational tuning process. However, in each of the benchmarks we consider large performance improvements are gained by changing the algorithm itself, which is not an automatic process. While the performance bottlenecks such too many barriers in BP1.0 and high data fetch to flop ratio in BP3.0 and BP3.5 are often not hard to identify, they can be troublesome to remove. We use a mixture of standard optimization strategies such as splitting the data between registers and shared memory and more advanced strategies, which resulted in redesigning the algorithm to make it more well-suited for fine-grain parallelism.

An obvious question remaining is how much more performance can be obtained in each of these benchmarks. In this paper we provide contributions towards the answer. Unlike \citep{Abdelfattah2016}, we consider the limitation of shared memory bandwidth in our approach. By no mean our model is exhaustive and our future work involves introducing a more advanced performance model that accounts for register use and other factors but retains the similarity of the original model. The source code for all kernels and benchmarks considered in this paper is publicly available at: \url{http://github.com/tcew/CEED}.

\section{Acknowledgements}
This research was supported in part by the Exascale Computing Project (17-SC-20-SC), a collaborative effort of two U.S. Department of Energy organizations (Office of Science and the National Nuclear Security Administration) responsible for the planning and preparation of a capable exascale ecosystem, including software, applications, hardware, advanced system engineering, and early testbed platforms, in support of the nations exascale computing imperative.

In addition, the authors would like to kindly acknowledge Advance Research Computing at Virginia Tech for providing readily accessible computational resources. Finally, this research was supported in part by the John K. Costain Faculty Chair in Science at Virginia Tech.

\bibliography{BP3paper}
\end{document}